\documentclass[aps,twocolumn,prb,amsmath,amssymb,showpacs]{revtex4}

\usepackage{graphicx,epsfig}
\usepackage{wasysym}

\begin{document}
\title{One-channel conductor coupled to a quantum of resistance:\\
exact finite-frequency conductance and noise}
\author{Redouane Zamoum}
\author{Adeline Cr\'epieux}
\affiliation{Centre de Physique Th\'eorique, Aix-Marseille University,\\
163 avenue de Luminy, FR-13288 Marseille, France}
\author{In\`es Safi}
\affiliation{Laboratoire de Physique des Solides, Universit\'e Paris-Sud,\\
B\^at. 510, FR-91405 0rsay Cedex, France\\}

\begin{abstract}
We consider a one-channel coherent conductor with a good transmission embedded into an Ohmic environment, impedance of which is equal to the quantum of resistance $R_q=h/e^2$ below the $RC$ frequency. This choice is motivated by the mapping of this problem to a Tomonaga-Luttinger liquid with one impurity, the interaction parameter of witch corresponds to the specific value $K=1/2$, allowing for a refermionization procedure. The ``new'' fermions have an energy-dependent transmission amplitude which incorporates the strong correlation effects and yields the exact dc current and zero-frequency noise through expressions similar to those of the scattering approach. We recall and discuss these results for our present purpose. Then we compute, for the first time, the finite-frequency differential conductance and the finite-frequency non-symmetrized noise. Contrary to intuitive expectation, both can not be expressed  within the scattering approach for the new fermions, even though they are still determined by the transmission amplitude. Even more, the finite-frequency conductance obeys an exact relation in terms of the dc current which is similar to that derived perturbatively with respect to weak tunneling within the Tien-Gordon theory,  and extended recently to arbitrary strongly interacting systems coupled eventually to an environment or/and with a fractional charge. We also show that the emission excess noise vanishes exactly above $eV$, even though the underlying Tomonaga-Luttinger liquid model corresponds to a many-body correlated system. Our results apply for all ranges of temperatures, voltages, and frequencies below the $RC$ frequency, and they allow us to explore fully the quantum regime.
\end{abstract}

\maketitle

\section{Introduction}

Laws of electrical circuits are drastically modified when mesoscopic systems are incorporated. A coherent conductor embedded into a circuit sees  the imposed voltage by the generator reduced  by a fluctuating voltage associated with the impedance of the surrounding electromagnetic environment. This gives rise to a current reduction with a pronounced non-linearity, called zero-bias anomaly (ZBA): transfer of electrons becomes inelastic as they exchange photons with the electromagnetic environment. This phenomena, called the dynamical Coulomb blockade (DCB), has attracted a tremendous interest both theoretically and experimentally.\cite{DCB} Nevertheless, most of the works  have been initially developed by focusing on the limit of a  weakly transmitting conductor, i.e., the tunneling regime.\cite{note1} More recently, interest in the regime of few well-transmitting channels has emerged. On one hand, reducing the number of channels makes more apparent the effect of the circuit, as many channels could play as well the role of an out-of-equilibrium environment. On the other hand, highly transmitting conductors raise two interesting questions. First, whether the charge fluctuations wash out the DCB, and second, whether the reduction in the current is related to shot noise in the absence of the environment. Indeed, this intuitive and attractive relation had been proposed through a perturbative computation with respect to a very weak impedance, and has been checked experimentally.\cite{golubev01,levyyeyati01,altimiras07} 
Nevertheless, its validity domain is restricted to high enough energies, as logarithmic divergences arise: one needs to go beyond the weak feedback action. This was particularly the case for an Ohmic environment, i.e., having an impedance $Z(\omega)=R$ at frequencies $\omega<\omega_{RC}=1/RC$ (where $R$ is the resistance and $C$ the capacitance):  this situation has not only relevance to realistic experiments but also a fundamental interest, being related to the investigation of electronic interactions. The logarithmic divergences have been resumed using a renormalization group (RG) scheme by Kindermann and Nazarov,\cite{kinderman03}  dealing again with $R\ll R_q$, where $R_q=h/e^2$ is the quantum of resistance. It has been nevertheless possible, in a simultaneous and independent work, to deal for the first time with an arbitrary value of $R$: one of the authors, with Saleur, \cite{safi04} has shown that the problem of a short-coherent conductor in series with a resistance $R$ is equivalent to the impurity problem in a Tomonaga-Luttinger liquid\cite{tomonaga50,luttinger63} (TLL) with an interaction parameter:
\begin{equation}\label{key}
K=(1+R/R_q)^{-1}~.
\end{equation}
When specified to the limit of small $R$, this mapping had led to recover the same results as those by Kindermann and Nazarov. These findings answer the two questions addressed above. First, the DCB still persists at good transmission, showing up below an energy scale $eV_B$ (depending in a non-universal way on $R$ and on transmission): it corresponds to the crossover voltage between the so-called weak backscattering (WBS) regime at high energy and strong backscattering (SBS) regime at low energy.\cite{note3} Secondly, the reduction in the current is related to the noise in the presence of the environment, and not to the noise of the isolated conductor as stated before (see Eq.~(\ref{crucial})). 

Recent pioneering experiments by Pierre's group\cite{pierre11} -- where the strong feedback of an arbitrary impedance on a one-channel edge state has been investigated for the first time -- has shown satisfactory agreement with the theoretical predictions of Refs.~\onlinecite{kinderman03} and \onlinecite{safi04}. Interestingly, even though one has to take strictly the limit of one channel in Ref.~\onlinecite{safi04}, the equivalence to a TLL thus established seems to extend to many channels  when they are treated in a mean-field framework, provided one renormalizes the parameter $K$ by including the resistance of the channels, and scales the voltage appropriately, as one can infer from a more recent study.\cite{golubev05}

By mapping a one-channel conductor in series with an Ohmic environment to a TLL, the parameter $K$ of which can be controlled by tuning $R$ (see Eq.~(\ref{key})), one gets as well a promising alternative to test the theoretical predictions for the impurity problem in a TLL. Indeed, satisfactory experimental evidence has been lacking, even though partial success has been claimed to be obtained. In particular, as noticed in Ref.~\onlinecite{safi04}, the case of an environmental resistance  equal to the quantum of resistance, $R=R_q$, corresponds to a TLL with an interaction parameter $K=1/2$. In that case, the impurity problem in a TLL can be solved exactly in a more transparent  way compared to other values of $K$ (see Ref.~\onlinecite{fendley95} for instance). Such an apparent simplicity is due to the introduction of new chiral fermions which arise from a mathematical construction without any physical entity, and incorporate non-trivial strong correlation effects. The non-perturbative investigation has a fundamental interest: it explores crucial issues not reached through perturbative approaches. Nevertheless, the situation with $K=1/2$ has never been achieved experimentally. Indeed, the ideal candidate to investigate the TLL's behavior is usually provided by edge states with a constriction in the fractional quantum Hall effect (FQHE). However, as $K$ plays the role of the fractional filling factor $\nu$, and as  the description of the edges in terms of one-channel chiral TLL model is valid only for simple values: $\nu=K=1/(2n+1)$ with $n$ an integer, it has not been possible to achieve a chiral TLL with $K=1/2$. Other potential candidates are provided by quantum wires and carbon nanotubes. However three main difficulties arise in those systems: (i) achieving only one backscattering center; (ii) tuning the interaction parameter $K$; and (iii) taking into account the connection to reservoirs and finite size effects, where only a perturbative treatment of the impurity has been possible.\cite{safi96,furusaki96,dolcini05,lebedev05,guigou07} Thus, a coherent one-channel conductor connected to a quantum of resistance $R=R_q$, as achieved recently,\cite{pierre11} offers an unique opportunity to explore the properties of a TLL with an impurity at $K=1/2$. 

The specific value, $K=1/2$, is exciting as it allows us to obtain handy analytic expressions for dc current and the zero-frequency noise. Beyond the stationary regime, time-dependent transport probes even more the dynamics and tests in a precise way the underlying model. 

This is the case of finite-frequency (FF) noise. In view of the mapping in Ref.~\onlinecite{safi04}, one can use previous results obtained in the FQHE. On the one hand, as far as the FF symmetrized noise is concerned, Chamon {\it et~al.}\cite{chamon96} have computed it perturbatively with respect to the impurity strength, and non-perturbatively at the specific value $K=1/2$. On the other hand, within the framework of the exact solution of Ref.~\onlinecite{fendley95}, Lesage and Saleur\cite{lesage97} have obtained only its behavior for low frequencies or close to the ``Josephson'' singularity $e^*V$, where $V$ denotes the voltage and $e^*=Ke$.\cite{noteq} Notice that both approaches disagree, apart from the particular value $K=1/2$. 

Nevertheless, it has been possible  to measure experimentally the FF non-symmetrized noise\cite{NSnoise}, which is more interesting to explore. It can be inferred from  Ref.~\onlinecite{bena07} dealing with edge states at simple filling factors,  both in the WBS and SBS regimes: the same results apply to a coherent conductor with a good bare transmission $\tau_0$ connected to an arbitrary resistance  at high or low enough energies, as well as for a low transmission.\cite{note2} Our aim is to go beyond this perturbative computation, offering a full description extending over all energy ranges below $\omega_{c}=\mathrm{min}\{\omega_F,\omega_{RC}\}$, where $\omega_F$ is the frequency cutoff associated to the fermionic degrees of freedom in the conductor. This is precisely possible at $R=R_q$. Thus, our study gives the first non-perturbative results for the FF non-symmetrized noise, without assuming neither weak resistance, nor high or low enough energies associated to the WBS nor SBS regimes. In addition, this offers a benchmark for other values of the resistance. 

One of the key steps within the refermionization procedure is that the chiral independent fermions have now an energy-dependent transmission amplitude $t(\omega)$ (see Eq.~(\ref{tomega})) which encodes the non-trivial many-body correlations.  Indeed, the associated transmission coefficient $\mathcal{T}(\omega)=|t(\omega)|^2$ has nothing to do with the effective transmission obtained for $K\approx 1$ by the RG approach,\cite{kinderman03,yue94} and can not be obtained adiabatically from that in the absence of interactions, $\tau_0$. It has been widely accepted that transport properties of the system can be obtained within the scattering approach using $t(\omega)$. This was shown to be valid both for the dc current, as well as for the full counting statistics (FCS).\cite{kinderman05} Surprisingly, our study shows that the scattering approach fails when one deals with time-dependent transport: the FF noise can {still} be expressed in terms of $t(\omega)$, nevertheless it obeys a different relation. The same feature occurs for another quantity: the out-of-equilibrium FF dissipative  conductance $\mathrm{Re}[G(V,\omega)]$ which depends on the applied dc voltage and the frequency of the superimposed modulation, in addition to its implicit dependence on temperature.

This last quantity has started to be studied only recently in few correlated systems, where its interest has been shown: quantum wires with an impurity\cite{safi08} and Kondo problem.\cite{simon11} This has become possible owing to a crucial result: $\mathrm{Re}[G(V, \omega)]$ is given by an out-of-equilibrium Kubo type formula, i.e., by the asymmetry between the emission and absorption FF noise.\cite{gavish03,safi09,safijoyez} In this present and first non-perturbative investigation, we show that $\mathrm{Re}[G(V, \omega)]$ does not fit with its expression obtained within the scattering approach. Even more, it obeys exactly a surprising relation, Eq.~(\ref{eq_G_I}), which determines it fully from the dc current, in a way similar to that obtained in tunnel junctions within the Tien-Gordon theory, but with a renormalized charge here.\cite{tunnel,tucker} Recently, such a relation has been generalized to full extent, including arbitrary interactions: it has been shown to be universal to lowest order with respect to a local or spatially extended tunneling at arbitrary dimension, as well as with respect to local or spatially extended backscattering if one-dimensional systems are considered.\cite{safi10,safi11} 

Notice that our results will depend on  parameters such as the frequency cutoff $\omega_c$, an effective transmission we call $\tau$, and the voltage scale $V_B$, which is related in a non-universal way to $\omega_c$ and $\tau$. Nevertheless, we will show that the scaling laws, known to be obeyed by the dc current, can be extended to the FF conductance and FF noise. More precisely, all these quantities depend only on $V/V_B$, $k_BT/eV_B$ and $\hbar\omega/eV_B$, where $T$ denotes the temperature.

\begin{table}
\begin{center}
\begin{tabular}{|l|l|} 
\hline
DCB & dynamical Coulomb blockade\\
FCS & full counting statistics\\
FF & finite-frequency\\
FQHE & fractional quantum Hall effect\\
RG & renormalization group\\
TLL & Tomonaga Luttinger liquid\\
SBS & strong backscattering\\
WBS & weak backscattering\\
ZBA & zero-bias anomaly\\
ZF & zero-frequency \\
\hline
\end{tabular}
\caption{Synthesis of the abbreviations.}
\label{abbreviations}
\end{center}
\end{table}

The paper is organized as follows: In Sec.~II, we review the mapping to an impurity problem in the TLL model, which is used to take the Ohmic environment into account, and some of its  general consequences on dc transport for arbitrary values of the environmental resistance.\cite{safi04} Next, in Sec.~III, we present our calculations performed in the case $R=R_q$, and give the formal expressions of the current, noise and conductance in term of transmission amplitude. In Sec.~IV, we discuss in details the dc regime (dc current, differential conductance, and zero-frequency (ZF) noise) for which we recover known results.\cite{guinea85,weiss91,kane92bis,fendley95bis} In Sec.~V, we explicit new results concerning the time-dependent transport: FF conductance and FF non-symmetrized noise, obtained in all energy ranges, and explore them in the quantum regime. We finally conclude in Sec.~VI. Notice that Tab.~\ref{abbreviations} refers to the adopted abbreviations and Tab.~\ref{parameters} gives a synthesis of the parameters and constants that are used.


\section{Model and outlook}

\begin{figure}[!h]
\begin{center}
\includegraphics[width=6cm]{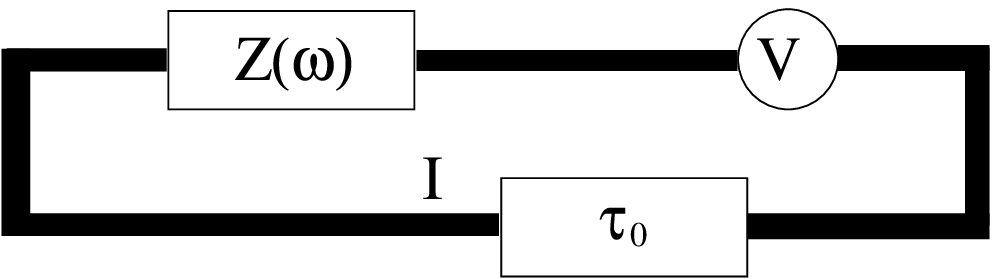}
\includegraphics[width=7cm]{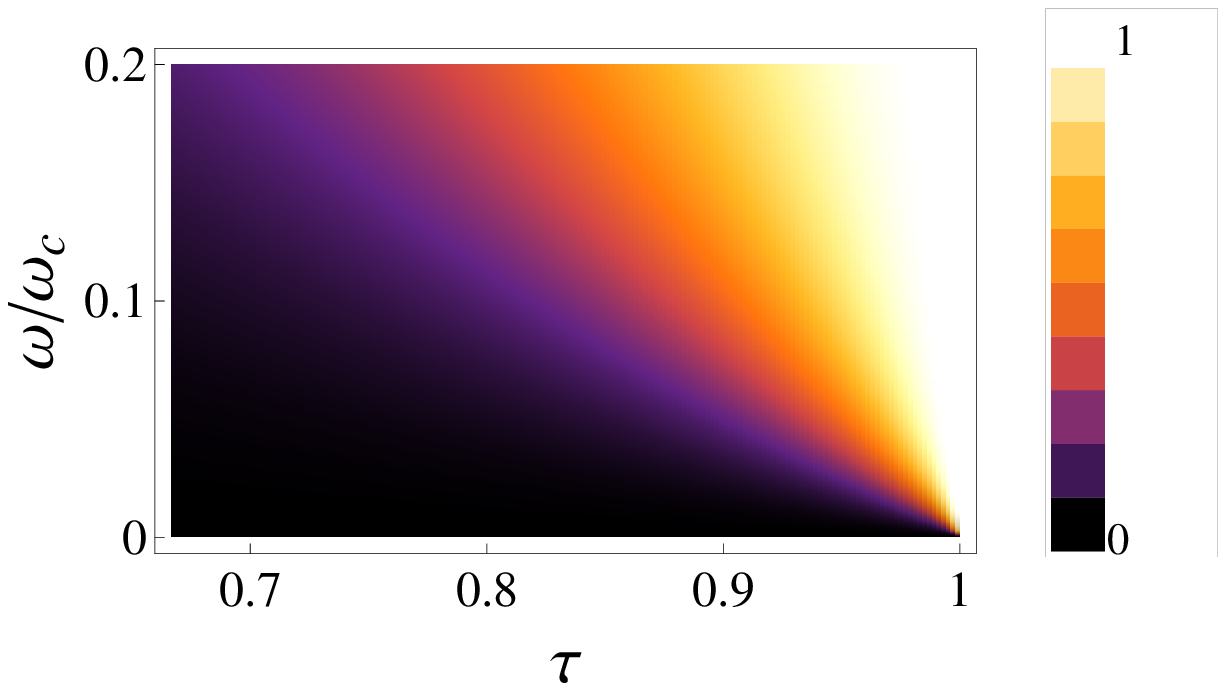}
\caption{Upper panel: Schematic representation of a one-channel conductor with bare transmission $\tau_0$ embedded in an electric circuit. Lower panel: Profile of the frequency dependent transmission coefficient $\mathcal{T}$ of the equivalent system as a function of frequency $\omega/\omega_c$ and effective transmission $\tau$ (see Eq.~(\ref{eq_T})).
\label{figure1}}
\end{center} 
\end{figure}

We consider a one-channel coherent conductor with bare transmission $\tau_0$, in series with a dissipative environment, the effective capacitance $C$ of which includes implicitly that of the conductor, thus whose impedance reads as $Z(\omega)=R/(1+i\omega RC)$. However, we will restrict to energies below $\omega_{RC}=1/RC$ in the following, thus one can approximate:
 \begin{equation}
Z(\omega)\approx R~, ~~\text{for} ~~ \omega<\omega_{RC}.
\end{equation}
We denote the voltage imposed by the generator by $V$, and the current through the circuit by $I$ (see upper panel in Fig.~\ref{figure1}). The voltage drop across the conductor is generally different from $V$. In the trivial limit of $\tau_0=1$, it is given simply by $KV$, where $K$ is defined by Eq.~(\ref{key}), due to the resistance in series of the perfect conductor and the resistance $R$ of the Ohmic environment. Whenever $\tau_0<1$, this is no more the case, apart from the perturbative regime with respect to $1-\tau_0$, when $\tau_0$ is close to one.

From now to the end of this section, we review the result of Ref.~\onlinecite{safi04}. The one-channel conductor in series with a resistance has been shown\cite{safi04} to be equivalent to an impurity problem in a TLL, the parameter $K$ of which is given by Eq.~(\ref{key}). Accordingly, one can use the TLL model in the presence of backscattering with amplitude $v_B$, the Hamiltonian of which reads as:
\begin{eqnarray}\label{hamilton}
\mathcal{H}&=&\mathcal{H}_0+\frac{\hbar\omega_F v_B}{4\pi\sqrt{\pi}}e^{i\phi(t)-ieKVt/\hbar}+h.c.~,
\end{eqnarray}
where $\mathcal{H}_0$ is the TLL Hamiltonian (for simplicity, we do not include spin degrees of freedom). The bosonic field $\phi$ coincides with the charge that is transferred through the conductor: $e\phi(t)=Q(t)$. This bosonized Hamiltonian is an effective one valid at energies below a typical energy $\hbar\omega_F$. In combination with the condition $\omega<\omega_{RC}$ for the mapping to hold, we denote the effective frequency cutoff by $\omega_c=\mathrm{min}\{\omega_F,\omega_{RC}\}$.\cite{notecutoff} In Eq.~(\ref {hamilton}), the parameter $K$ in $exp(-ieKVt/\hbar)$ results from the dc conductance of the ballistic conductor (as can be inferred for instance from  Ref.~\onlinecite{dolcini05}). Moreover, the effective backscattering amplitude $v_B$ is not related universally to the bare transmission $\tau_0$ in the absence of the environment. Indeed one can not, in general, express the parameters of a strongly  correlated system in terms of those without interactions, lacking correspondence between them.\cite{note10} Nevertheless, we introduce an effective transmission $\tau=1/(1+v_B^2)$ keeping in mind that $\tau$ does not have to coincide with $\tau_0$. Of course, when $\tau_0=1$ one has no backscattering, and $\tau=1$ too. Also, we restrict $v_B$ to be weak enough -- thus $\tau$ close to one -- in order to write the backscattering Hamiltonian in its bosonized form on the r.h.s. of Eq.~(\ref {hamilton}). It is reasonable, though not well established,  that this would correspond to $\tau_0$ close to one as well. It is however possible that the Hamiltonian in Eq.~(\ref{hamilton}) could be extended beyond that restriction.\cite{egger98} 

\begin{table}
\begin{center}
\begin{tabular}{|l|l|} 
\hline
$R$, $C$, $Z$&resistance, capacitance, \\
&and impedance of the environment\\ 
$K=(1+R/R_q)^{-1}$&interaction parameter of the TLL\\
$R_q=h/e^2$&quantum of resistance\\
$G_q=e^2/h$&quantum of conductance\\
$v_F$&Fermi velocity\\
$a$&distance cutoff of the TLL\\
$\omega_F=v_F/a$&frequency cutoff of the TLL\\
$\omega_{RC}=1/RC$&frequency of the RC circuit\\
$\omega_c=\min\{\omega_F,\omega_{RC}\}$&frequency cutoff of our model\\
$\tau_0$&bare transmission\\
$\tau$&effective transmission\\
$v_B=\sqrt{(1-\tau)/\tau}$&backscattering amplitude\\
$eV_B=2\hbar\omega_cv_B^2$&energy crossover between WBS and SBS\\
\hline
\end{tabular}
\caption{Synthesis of parameters and constants.}
\label{parameters}
\end{center}
\end{table}

The important effective parameter is indeed the scaling voltage $V_B$,\cite{note3} which appears in the Bethe-Ansatz solution as non-universal.\cite{fendley95} It can be roughly related to the backscattering amplitude through $eV_B\simeq \hbar\omega_c v_B^{1/(1-K)}$. Indeed $V_B$ characterizes the crossover between the WBS and SBS regimes. At energies high enough compared to  $eV_B$, one can use the perturbative RG analysis with respect to a weak impurity by Kane and Fisher.\cite{kane92} In the opposite limit (i.e., at energies $\ll eV_B$) the wire is cut into two pieces, with weak tunneling between them. Thus, even though one starts from a weak impurity, lowering the energy drives the system from the WBS behavior, where the conductance is slightly reduced, into the SBS regime where the conductance is suppressed. This means that the DCB still takes place even when the effective transmission $\tau$ is close to one,\cite{safi04} but below an effective charging energy  $eV_B$. In particular, at energies much smaller than $eV_B$, thus in the SBS regime, it has been shown that one recovers the $P(E)$ theory which is rather obtained starting from a very weak transmission.\cite{DCB} Even more, it is possible to obtain non-perturbative results not only with respect to $R$, but also to $1-\tau$, and to describe the whole regime of energies below $\omega_{c}$. Such an achievement was made possible by exploiting the Bethe-Ansatz exact solutions of Fendley {\it et~al.}\cite{fendley95} for the impurity problem in a TLL. Within the Bethe-Ansatz solution, the FCS can be computed exactly as well. In particular, using these results for the current and noise, it was possible to derive a crucial and exact relation between the derivative of the differential conductance $G(V,\omega=0)=dI(V)/dV$ and the differential ZF noise $S(V,\omega=0)$ at zero temperature:
\begin{eqnarray}\label{crucial}
R_q|eV|\;dG(V,\omega=0)=2R\;dS(V,\omega=0)~.
\end{eqnarray}
Let us remark that in the limit of small $R\ll R_q$, Eq.~(\ref{crucial}) fits perfectly with the RG equation obtained by Kindermann and Nazarov.\cite{kinderman03}  Within the Bethe-Ansatz solution, the FCS can be computed exactly as well. Using its derivation at zero temperature,\cite{saleur01} the mapping permits to extend Eq.~(\ref{crucial}) to higher cumulants. It has also motivated partly the recent investigations of the FCS at finite temperature.\cite{kinderman05,komnik06,komnik07,herzog08,komnik}

More recently, in an interesting work, Golubev {\it et~al.}\cite{golubev05} have studied a mesoscopic interacting multi-channel conductor connected to an Ohmic environment, using a method based on Keldysh action.  They have confirmed the results presented above (of Ref.~\onlinecite{safi04}) in the perturbative regime they restrict to (with respect to $1-\tau$),  which corresponds to high energies in the WBS domain.\cite{notesurtau} Even more, when treating many channels in a mean-field approach, the agreement holds too, provided the TLL's parameter in Eq.~(\ref{key}) incorporates the resistance of the conductor. This shows that the mapping can be extended to many channels, as well as its consequences discussed above (at least at high enough energy), and motivates further the computation done in this work.

In the following, we will consider the situation where the resistance is equal to the quantum of resistance: $R=R_q$. In that case, the TLL parameter is $K=1/2$ (see Eq.~(\ref{key})), and Eq.~(\ref{hamilton}) is exactly solvable through a refermionization procedure,\cite{guinea85,chamon96,matveev95,furusaki97,vondelft98,aristov09} thus one can perform a non-perturbative analysis of  transport properties. The refermionization introduces new independent chiral fermions with a frequency dependent transmission amplitude:
\begin{eqnarray}\label{tomega}
t(\omega)=\frac{\omega}{\omega+ieV_B/2\hbar}~,
\end{eqnarray}
thus a transmission coefficient:
\begin{eqnarray}\label{eq_T}
\mathcal{T}(\omega)=|t(\omega)|^2=\frac{4\hbar^2\omega^2}{4\hbar^2\omega^2+e^2V_B^2}=\frac{\tau^2\omega^2}{\tau^2\omega^2+(1-\tau)^2\omega_c^2}~,
\end{eqnarray} 
where $eV_B=2\hbar\omega_cv_B^2$ is the energy crossover between the WBS and SBS regimes. The profile of  $\mathcal{T}(\omega)$  is shown in the lower panel of Fig.~\ref{figure1}. Obviously, $\mathcal{T}(\omega)$ is identical to $1$ (perfect effective transmission) when $\tau=1$ whatever the frequency is. As we will show later on,  $\mathcal{T}(\omega=eV/\hbar)$ yields the non-linear differential conductance $G(V,\omega=0)$ at zero temperature. In accordance with the features concerning the crossover from WBS to SBS, one sees that as soon as $\tau$ deviates from one, $\mathcal{T}(\omega)$ decreases quickly to zero at low frequency compared to $eV_B/\hbar$. From Eq.~(\ref{eq_T}), notice that  $\mathcal{T}(\omega=v_B\omega_c)$ coincides with the effective transmission $\tau$. 


\section{Results}

In this section, we present the formal results for the dc current, the non-symmetrized noise, and the differential conductance.

We first calculate the dc current, defined as the average of the time derivative of the charge which is transferred through the conductor: $I(V)=\langle \hat{I}(t)\rangle=\langle \dot Q(t)\rangle$, where $\hat{I}$ is the current operator. Since we consider a dc applied voltage, $I(V)$ is time-independent. The details of the calculation are presented in Appendix~\ref{app_current}. 
We obtain:\cite{note4}
\begin{eqnarray}\label{eq_current}
I(V)&=&\frac{e}{4\pi}\int_{-\infty}^{\infty}d\omega\mathcal{T}(\omega)\nonumber\\
&&\times\left[f\left(\hbar\omega-eV/2\right)-f\left(\hbar\omega+eV/2\right)\right]~,
\end{eqnarray}
which corresponds to the Landauer formulation of the current, where the Fermi-Dirac distribution function is given by $f(\hbar\omega)=[1+\exp(\hbar\omega/k_BT)]^{-1}$. The density of states multiplied by the velocity is a constant, equal to $1/2\pi$, because the energy spectrum of the new fermions is linear too. Eq.~(\ref{eq_current}) describes the behavior of the dc current over all voltage and temperature ranges, starting from the WBS regime down to the SBS regime. We have to recall that  the model describes a strongly correlated system where interactions cannot be treated by any mean-field approach, and that the transmission amplitude incorporates their non-trivial effects. Notice that in Eq.~(\ref{eq_current}) we have made the choice of extending the limits of integration to plus and minus infinity. Strictly speaking, these limits should be $-\omega_c$ and $\omega_c$, however we have checked that the correction terms are negligible. In the following, a similar choice will be made in all integrals over frequencies.

Next, we calculate the FF non-symmetrized noise defined as the Fourier transform of the current fluctuations:
\begin{eqnarray}
S(V,\omega)=\int_{-\infty}^{\infty}dte^{i\omega t}\langle \delta \hat{I}(0)\delta \hat{I}(t)\rangle~,
\end{eqnarray}
where $\delta \hat{I}(t)=\hat{I}(t)-\langle \hat{I}\rangle$. We obtain the following result (see Appendix~\ref{app_noise} for details):
\begin{widetext}
\begin{eqnarray}\label{eq_noise}
S(V,\omega)&=&\frac{e^2}{4\pi}\sum_\pm\int_{-\infty}^{\infty}d\omega'\bigg(\Big[\mathcal{T}(\omega')\mathcal{T}(\omega+\omega')+|t(\omega')-t(\omega+\omega')|^2/4\Big][1-f(\hbar\omega'\pm eV/2)]f(\hbar\omega'+\hbar\omega\pm eV/2)\nonumber\\
&&+ \Big[\mathcal{T}(\omega')-\mathcal{T}(\omega')\mathcal{T}(\omega+\omega')-|t(\omega')-t(\omega+\omega')|^2/4\Big][1-f(\hbar\omega'\pm eV/2)]f(\hbar\omega'+\hbar\omega\mp eV/2)\bigg)~.
\end{eqnarray}
\end{widetext}
A crucial and surprising observation is that this expression differs from that obtained within the scattering approach for a one-channel conductor with an energy-dependent transmission\cite{martin92,buttiker92,yang92,blanter00} (see Appendix~\ref{app_comp} for more details on that comparison), even when applied to chiral fermions.\cite{lev12}

It is interesting to cast Eq.~(\ref{eq_noise}) under the alternative form:
\begin{eqnarray}\label{eq_noise_bis}
&&S(V,\omega)=\frac{e}{2}\sum_\pm N(\hbar\omega\pm eV)[I(\pm V)+I(2\hbar\omega/e\pm V)]
\nonumber\\
&&+\frac{e^2}{4\pi}\sum_\pm [N(\hbar\omega\pm eV)-N(\hbar\omega)]
\nonumber\\
&&\times\int_{-\infty}^{\infty}d\omega'
\left[\mathcal{T}(\omega')\mathcal{T}(\omega+\omega')+\frac{|t(\omega')-t(\omega+\omega')|^2}{4}\right]\nonumber\\
&&\times[f(\hbar\omega+\hbar\omega'\pm eV/2)-f(\hbar\omega'\mp eV/2)]~,
\end{eqnarray}
where $N(\hbar\omega)=[\exp(\hbar\omega/k_BT)-1]^{-1}$ is the Bose-Einstein distribution function. Notice that in the trivial limit $\tau=\tau_0=1$ (i.e., perfect transmission),  the voltage drop across the mesoscopic conductor is $V/2$ and the FF noise reduces to $S(V,\omega)=\hbar\omega N(\hbar\omega)G_q$ whatever the temperature is, where $G_q=e^2/h$ is the quantum of conductance. An interesting point is that, in the r.h.s. of  Eq.~(\ref{eq_noise_bis}), both distribution functions $f$ for fermions and $N$ for bosons (of the electromagnetic environment, thus electron-hole excitations) are involved, which explains the fact that the effective voltage is not the same in their arguments: one has $\hbar\omega\pm eV/2$ for the function $f$, and $\hbar\omega\pm eV$ for the function $N$.

From Eq.~(\ref{eq_noise}), we deduce immediately the ZF noise:
\begin{eqnarray}\label{eq_zero_noise}
&&S(V,0)=\frac{e^2}{4\pi}\sum_\pm\int_{-\infty}^{\infty}d\omega'\nonumber\\
&&\times\Big(\mathcal{T}^2(\omega')[1-f(\hbar\omega'\pm eV/2)]f(\hbar\omega'\pm eV/2)\nonumber\\
&&+ \mathcal{T}(\omega')\left(1-\mathcal{T}(\omega')\right)[1-f(\hbar\omega'\pm eV/2)]f(\hbar\omega'\mp eV/2)\Big)~.\nonumber\\
\end{eqnarray}
Contrary to what occurs for the FF noise, one recovers, in the ZF limit, an expression in terms of the transmission coefficient $\mathcal{T}$ similar to that within the scattering theory.

Let us now calculate the FF conductance which is related to the non-symmetrized noise through the exact relation:\cite{safi08,safi09}
\begin{eqnarray}\label{def_G}
\mathrm{Re}[G(V,\omega)]=\frac{S(V,-\omega)-S(V,\omega)}{2\hbar\omega}~.
\end{eqnarray}

Reporting Eq.~(\ref{eq_noise}) in Eq.~(\ref{def_G}), all the terms which contain a product of transmission coefficients vanish, and we find simply:
\begin{eqnarray}\label{eq_G}
&&\mathrm{Re}[G(V,\omega)]=\frac{e^2}{4h\omega}\sum_\pm\int_{-\infty}^{\infty}d\omega'
\mathcal{T}(\omega')\nonumber\\
&&\times\Big[f(\hbar\omega'\pm eV/2)-f(\hbar\omega+\hbar\omega'\pm eV/2)\Big]~,
\end{eqnarray}
which obeys unexpectedly the exact relation:\cite{note5}
\begin{eqnarray}\label{eq_G_I}
\mathrm{Re}[G(V,\omega)]&=&\frac{e}{4\hbar\omega}\left[I\left(V+2\hbar\omega/e\right)-I\left(V-2\hbar\omega/e\right)\right]~.\nonumber\\
\end{eqnarray}

This is a central result of our paper: even though our computation is non-perturbative, one recovers a similar relation to that obtained within the perturbative Tien-Gordon theory, with a renormalized charge $e/2$.\cite{tunnel,tucker} Indeed, such a relation has been recently shown in Ref.~\onlinecite{safi10} to be generally valid for all strongly correlated systems at arbitrary dimensions, with possible coupling to an arbitrary electromagnetic environment, provided one requires the tunneling regime. In one-dimensional systems, the same work has shown the universality of the relation for one or many weak impurities in the WBS regime aside from its validity in the SBS regime or for tunneling barriers. Here, surprisingly, for a TLL with an impurity and $K=1/2$, or a conductor coupled to a quantum resistance, Eq.~(\ref{eq_G_I}) is valid for all voltage, temperature and frequency ranges (below $\hbar\omega_c$) with a renormalized charge $e/2$.

It is interesting as well to specify to frequencies much larger than the applied voltage $V$, where the FF conductance becomes voltage-independent,  thus reaches a linear regime. In that case, Eq.~(\ref{eq_G_I}) can be approximated simply, as in Ref.~\onlinecite{safi10}, by: 
\begin{eqnarray}
\mathrm{Re}[G(V\ll\hbar\omega/e,\omega)]\approx \frac{eI(2\hbar\omega/e)}{\hbar\omega}~.
\end{eqnarray}

In the two following sections, starting from the expressions derived here, we first recall some known results of the stationary regime (Sec.~IV) and next, we discuss in details the news results of the time-dependent regime (Sec.~V).


\section{Differential conductance and ZF noise}

In this section, we explicit in more details the dc transport in all the temperature and voltage ranges, starting from the low temperature regime ($k_BT\ll eV$) and ending with the high temperature regime ($k_BT\gg eV$).

{\it Low temperature behavior}. In the limit of strictly zero temperature, the integral over frequency in Eq.~(\ref{eq_current}) can be performed analytically and the dc current reads as:\cite{guinea85,weiss91,kane92bis}
\begin{eqnarray}\label{eq_current_T_0}
I(V)=\frac{G_qV_B}{2}\left[\frac{V}{V_B}-\;\mathrm{arctan}\left(\frac{V}{V_B}\right)\right]~.
\end{eqnarray}
The dc current is thus given by a product of $V_B$ multiplied by a function of $V/V_B$. This result has been used successfully to describe recent experimental data.\cite{pierre12} For perfect effective transmission, $\tau=1$ (i.e., $V_B=0$), we recover Ohm's law: $I(V)=V/(2R_q)$, whereas at $\tau< 1$, the dc current is reduced: DCB persists even though one starts from a good effective transmission. Even more, one can check, as shown in Ref.~\onlinecite{safi04}, that in the SBS regime, at $V\ll V_B$, one recovers a power law behavior controlled by the exponent $1+2R/R_q=3$ as in the $P(E)$ theory, and obtained rather for a weakly transmitting conductor.\cite{DCB}
For that, one expands Eq.~(\ref{eq_current_T_0}) with respect to $V/V_B$:
\begin{eqnarray}\label{ISBS}
I(V\ll V_B)&=&\frac{G_qV^3}{6V_B^2}~.
\end{eqnarray}
In the opposite limit of the WBS regime, i.e. at $V\gg V_B$, the reduction to the perfect current, called the backscattering current:
\begin{eqnarray}\label{back}
I_B(V)&=&\frac{G_qV}{2}-I(V)~,
\end{eqnarray}
can be expanded with respect to $V_B/V$:
\begin{eqnarray}\label{backexpansion}
I_B(V\gg V_B)&\simeq&\frac{G_qV_B}{2}\left(\frac{\pi}{2} -\frac{V_B}{V}\right)~.
\end{eqnarray}
Notice that the first term on the r.h.s. corresponds to that obtained within the perturbative computation,\cite{kane92bis} $I_B(V)\sim V^{2K-1}$, which becomes voltage independent at $K=1/2$.

\begin{figure}[h!]
\begin{center}
\includegraphics[width=4.2cm]{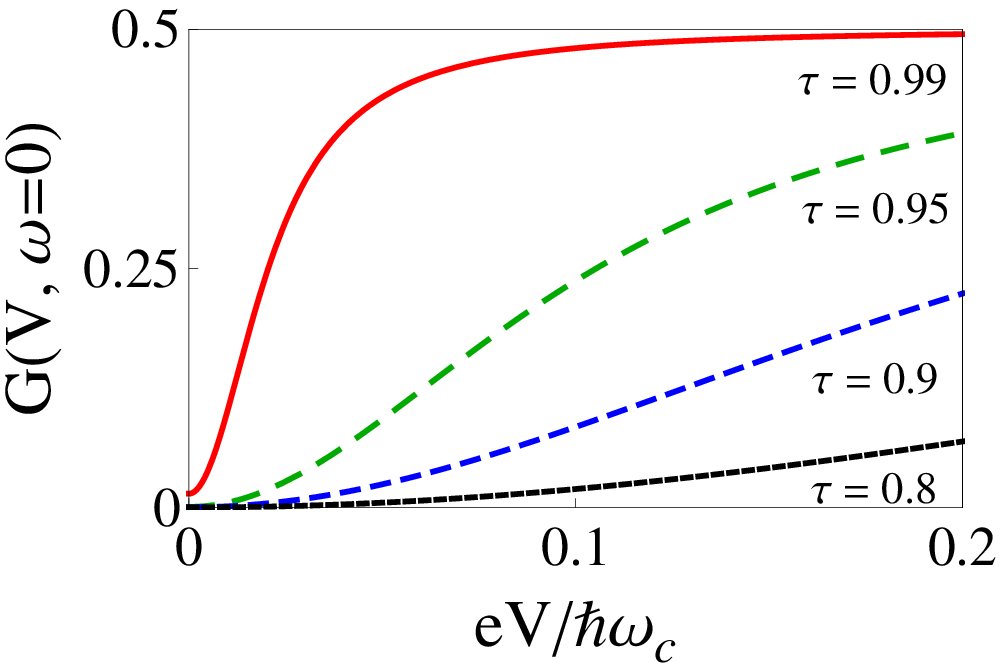}
\includegraphics[width=4.2cm]{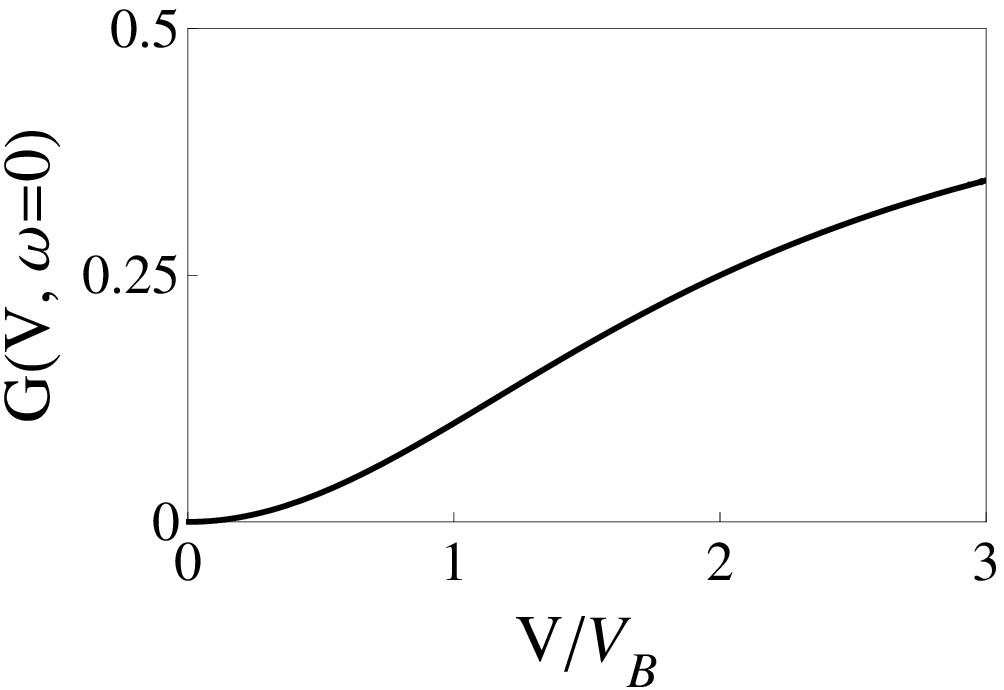}
\caption{Left panel: Differential conductance, in unit of $e^2/h$, as a function of $eV/\hbar\omega$, for different values of $\tau$, at $k_BT/\hbar\omega_c=0.001$. Since we have $eV_B/\hbar\omega_c=2(1-\tau)/\tau$, the corresponding values of $V_B$ are: $eV_B/\hbar\omega_c=0.02$ (red solid line), $eV_B/\hbar\omega_c=0.1$ (green dashed line), $eV_B/\hbar\omega_c=0.22$ (blue short dashed line), and $eV_B/\hbar\omega_c=0.5$ (black dotted line). Right panel: All the curves of the left graph scale to a single one when one considers the variation with $V/V_B$. We take $k_BT/eV_B=0.001$. \label{figure2}}
\end{center}
\end{figure}

The differential conductance can be obtained either by letting $\omega\rightarrow 0$ in Eq.~(\ref{eq_G}), or by differentiating the dc current of Eq.~(\ref{eq_current_T_0}). At zero temperature, it reads as:
\begin{eqnarray}\label{diff_cond_T_0}
G(V,\omega=0)=\frac{dI(V)}{dV}=\frac{G_q}{2}\left[1-\frac{V_B^2}{V^2+V_B^2}\right]~,
\end{eqnarray}
and obeys the relation:
\begin{eqnarray}
G(V,\omega=0)=\frac{G_q}{2}\mathcal{T}\left(\frac{eV}{2\hbar}\right)~,
\end{eqnarray} 
where the behavior of the transmission coefficient $\mathcal{T}$ is shown on the lower panel of Fig.~\ref{figure1}. Notice that, at $\tau=1$, the differential conductance becomes constant (linear): it is equal to $G_q/2$ as the ballistic conductor  is in series with a resistance $R=R_q$. At $\tau<1$, $G(V,\omega=0)$ can formally reach $G_q/2$ for $V\rightarrow \infty$, nevertheless it stays  below since $eV$ is limited by the cutoff $\hbar\omega_c$.

The left panel of Fig.~\ref{figure2} gives the differential conductance at low temperature, which shows a ZBA that is more pronounced when $\tau$ is reduced.  A surprising result is its large sensitivity with respect to small variations of $\tau$ in the vicinity of $1$ (see the red and green curves in Fig.~\ref{figure2} for example): it  is due both to the rapid variations of  $\mathcal{T}(\omega')$ at frequencies low compared to $\omega_c$ (when $\tau$ is close to $1$, see the lower panel of Fig.~\ref{figure1}) which are precisely those that give the dominant contribution to the integral over $\omega'$ in Eq.~(\ref{eq_G}).\cite{note_cond} We can see that the crossover $V_B$ varies rapidly in the vicinity of $\tau$ close to one, thus the SBS regime is reached much more rapidly when $\tau$ decreases. It is important to recall that all the curves scale to the same one when one scales the voltage by $V_B$, as shown in the right panel of Fig.~\ref{figure2}.

\begin{figure}[!ht]
\begin{center}
\includegraphics[width=6cm]{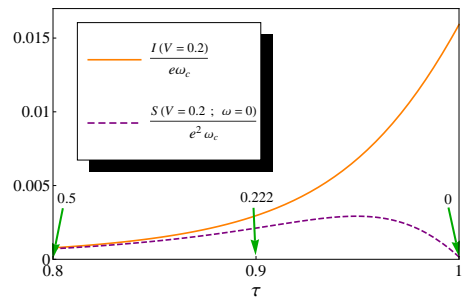}
\caption{dc current and ZF noise as a function of the effective transmission $\tau$, for $eV/\hbar\omega_c=0.2$, at a low temperature: $k_BT/\hbar\omega_c=0.001$. The arrows indicate the associated values of $eV_B/\hbar\omega_c$. 
 With a frequency cutoff of about $1$~THz, the value $0.01$ $e\omega_c$ on the axis corresponds to a current of about $10$~nA.}\label{figure3}
\end{center}
\end{figure}

At zero-temperature, the ZF noise of Eq.~(\ref{eq_zero_noise}) becomes:
\begin{eqnarray}
S(V,\omega=0)=\frac{G_qeV_B}{4}\left|\mathrm{arctan}\left(\frac{V}{V_B}\right)-\frac{VV_B}{V^2+V_B^2}\right|~.
\end{eqnarray}
As for the dc current, the ZF noise is  given by a product of $V_B$ multiplied by a function of $V/V_B$. One can check that the ZF noise obeys Eq.~(\ref{crucial}),
with $R=R_q$, which confirms that the ZBA is exactly linked to the ZF shot-noise in the presence of the electromagnetic environment.

Figure \ref{figure3} shows the dc current and the ZF noise as functions of $\tau$, at low temperatures compared to voltage, $k_BT\ll eV$: the current shows an increasing behavior, whereas the ZF noise is non monotonous. This is due to the fact that the current is expressed as the integral of the transmission coefficient, see Eq.~(\ref{eq_current}), and that the integral expression of the ZF noise of Eq.~(\ref{eq_zero_noise}) contains a term proportional to $\mathcal{T}(\omega')(1-\mathcal{T}(\omega'))$. For $V_B>V$ (i.e., $\tau<0.9$ using the parameters of Fig.~\ref{figure3}), the current and ZF noise curves converge to the same value. Thus, the Fano factor -- given by the ratio $S(V,\omega=0)/|I(V)|$ -- tends to $e$, the value obtained in the Poissonian regime with independent transfer events of charge $e$ trough the mesoscopic conductor. For $V_B\ll V$, the transfer events are no more independent but the noise is still close to zero because of the $(1-\mathcal{T}(\omega'))$ factor. In addition, the Fano factor -- defined this time by the ratio $S(V,\omega=0)/|I_B(V)|$, where $I_B$ is the backscattering current given by Eq.~(\ref{back}) -- is equal to $e^*=e/2$. In the FQHE at $\nu=K=1/(2n+1)$, the Fano factor would be given by  fractional charge $e^*=Ke$.  In the DCB context, this renormalization is related to the dc conductance $G_q/2$ without backscattering.

\begin{figure}[h!]
\begin{center}
\includegraphics[width=4.2cm]{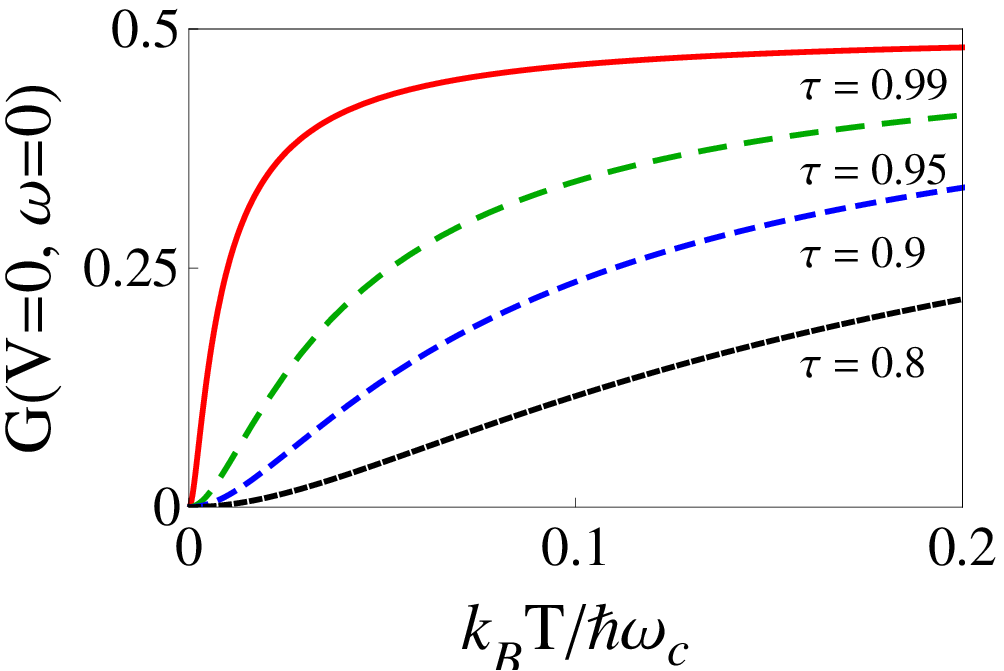}
\includegraphics[width=4.2cm]{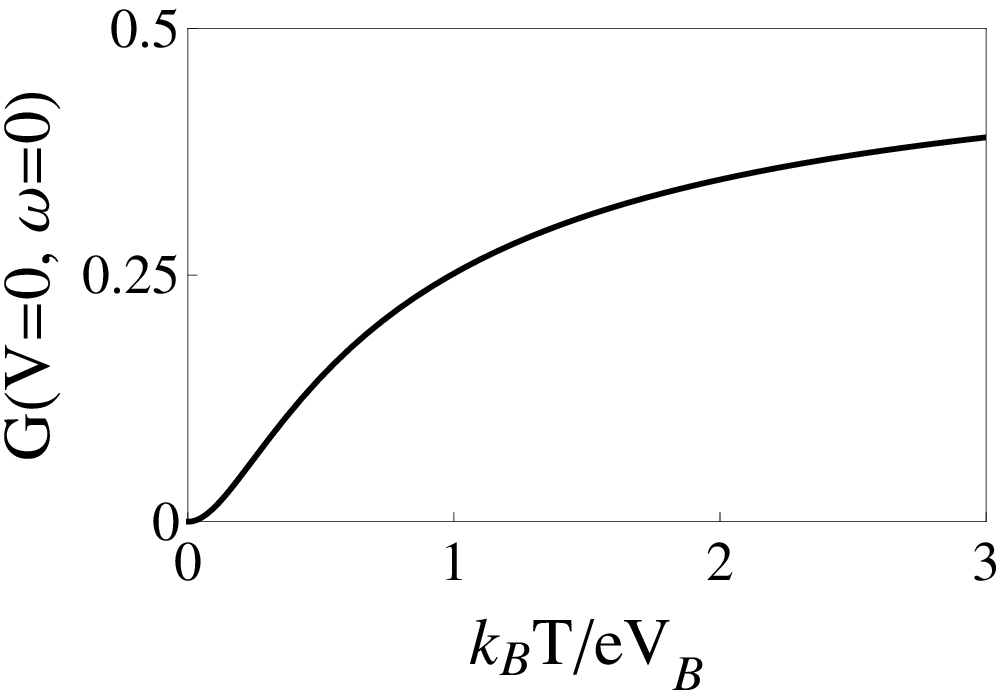}
\caption{Left panel: Differential conductance, in units of $e^2/h$, as a function of temperature $T$, in units of $\hbar\omega_c/k_B$, for different values of $\tau$, at $V=0$.  Right panel:  All the curves of the left graphic scale to a single one when one considers the variation with respect to $k_BT/eV_B$. 
\label{figure4}}
\end{center}
\end{figure}

\begin{figure}[!ht]
\begin{center}
\includegraphics[width=6cm]{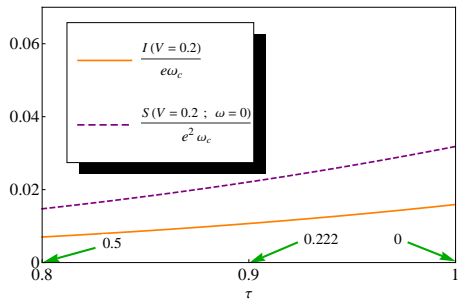}
\caption{dc current and ZF noise as a function of the effective transmission $\tau$, for $eV/\hbar\omega_c=0.2$, and $k_BT/\hbar\omega_c=0.2$. The arrows indicate the associated values of $eV_B/\hbar\omega_c$. \label{fig_zero_noise_2}}
\end{center}
\end{figure}

{\it Intermediate temperature behavior}. The left panel of Fig.~\ref{figure4} shows the differential conductance at zero voltage as a function of temperature for various values of $\tau$. When the temperature increases, the ZBA is suppressed, a behavior similar to what is obtained within the $P(E)$ theory.\cite{DCB,grivin90} Again, all these curves coincide when one considers their variation with respect to $k_BT/eV_B$ (see right panel of Fig.~\ref{figure4}). Fig.~\ref{fig_zero_noise_2} shows that the current and the ZF noise increase both monotonously with $\tau$, and that the noise is larger that the current due to thermal fluctuations. In the intermediate temperature regime $k_BT\approx eV$, the ZF noise has a totally different dependence on $\tau$ in comparison to its behavior at low temperature due again to an increasing contribution of the thermal noise. However, the total noise is not a direct superposition of shot noise and thermal noise, but results from a more complicated interplay between them.\cite{dolcini05,bena07,safi08}

{\it High temperature behavior}. In the equilibrium regime (i.e., for $k_BT\gg eV$), the dc current of Eq.~(\ref{eq_current}) becomes strictly linear in $V$, and the linear conductance takes the value:\cite{guinea85,weiss91,kane92bis,fendley95bis}
\begin{eqnarray}\label{diff_cond_V_0}
G&=&G(V=0,\omega=0)\nonumber\\
&=&\frac{G_q}{2}\left[1-\frac{eV_B}{4\pi k_BT}\Psi'\left(\frac{1}{2}+\frac{eV_B}{4\pi k_B T}\right)\right]~,
\end{eqnarray}
where $\Psi(x)=\Gamma'(x)/\Gamma(x)$, and $\Gamma$ is the Gamma function.  In particular, in the SBS, at $k_BT\ll eV_B$, one recovers the power law:
\begin{equation}\label{GSBS}
G=\frac{2\pi^2G_qk_B^2T^2}{3e^2V_B^2}~.
\end{equation}
This result is in accordance with the generic behavior of the TLL in the SBS regime,\cite{kane92,kane92bis} where one has $G\sim T^{2/K-2}$, as $K=1/2$ here. It is also similar to that obtained within the $P(E)$ theory in the tunneling regime, even though the conductor is well-transmitting here.\cite{safi04} Notice that this dependence on temperatures $k_BT\gg eV_B$ in Eq.~(\ref{GSBS}) is similar to that of the non-linear conductance on voltages $V\ll V_B$, as one can see from differentiating Eq.~(\ref{ISBS}). This confirms, as generally expected, that temperature and voltage play symmetric roles. More precisely, the differential conductance at both finite temperatures and voltages obeys a scaling law: $G\sim T^{\alpha}F(V/T)$,  where $F(x\ll 1)\simeq x^{\alpha}$ and $F(x\gg 1)\rightarrow$ constant value. This leads in particular to the same power law behavior with respect to $\mathrm{max}\{k_BT,eV\}$. While this is valid in the SBS regime as we have just shown, with $\alpha=2$, it turns out that this scaling behavior is violated in the opposite limit of WBS as we discuss in detail now. More precisely, one has to consider the backscattering conductance,  defined as the differential of the backscattering current in Eq.~(\ref{back}): $G_B=G_q/2-G$.  Using  Eq.~(\ref{diff_cond_V_0}) at $k_BT\gg eV_B$ yields:
 \begin{equation}\label{GWBS}
G_B=\frac{\pi G_qeV_B}{16k_BT}~,
\end{equation}
in accordance with the behavior $V_BT^{2 K-2}$ at arbitrary $K$. We see, however, that this dependence on temperature is different from that on voltage obtained  by expanding Eq.~(\ref{diff_cond_T_0}) in the WBS regime ($V\gg V_B$) (see also Eq.~(\ref{backexpansion})): 
\begin{equation}\label{GB}
G_B=\frac{G_qV_B^2}{2V^2}~.
\end{equation}
 As we have already commented concerning Eq.~(\ref{backexpansion}), for generic values of $K$, the lowest order expansion of the backscattering current is given by $I_B\sim V_BV^{2 K-1}$, leading to a differential conductance proportional to $(2 K-1)V_BV^{2 K-2}$, which cancels at $K=1/2$. This can be seen as well by differentiating the constant term on the r.h.s. of Eq.~(\ref{backexpansion}): thus one needs the next term, yielding to the first  non-zero contribution in Eq.~(\ref{GB}).


\section{FF conductance and FF noise}

In this section, we discuss the dynamical properties of our system by performing plot analysis of the formal expressions of the FF conductance $\mathrm{Re}[G(V,\omega)]$ and the FF non-symmetrized noise $S(V,\omega)$ obtained in Sec.~III. Both quantities depend not only of the voltage $V$ and frequency $\omega$, but also on temperature. More precisely, they can be cast into a scaling form, being a function of $V/V_B$, $\hbar\omega/eV_B$, and $k_BT/eV_B$. All the dependencies can be obtained exactly, provided these energies are below $\hbar\omega_c$, but we have to make restrictions on the number of curves presented. Here we will focus first on low temperatures compared to voltage and frequency, $k_BT\ll \{\hbar\omega,eV\}$, thus both out-of-equilibrium and quantum regime, allowing for $\hbar\omega$ to be of the order or greater than $eV$. Then we will consider intermediate and high temperatures compared to the voltage.

\begin{figure}[!h]
\begin{center}
\includegraphics[width=4.2cm]{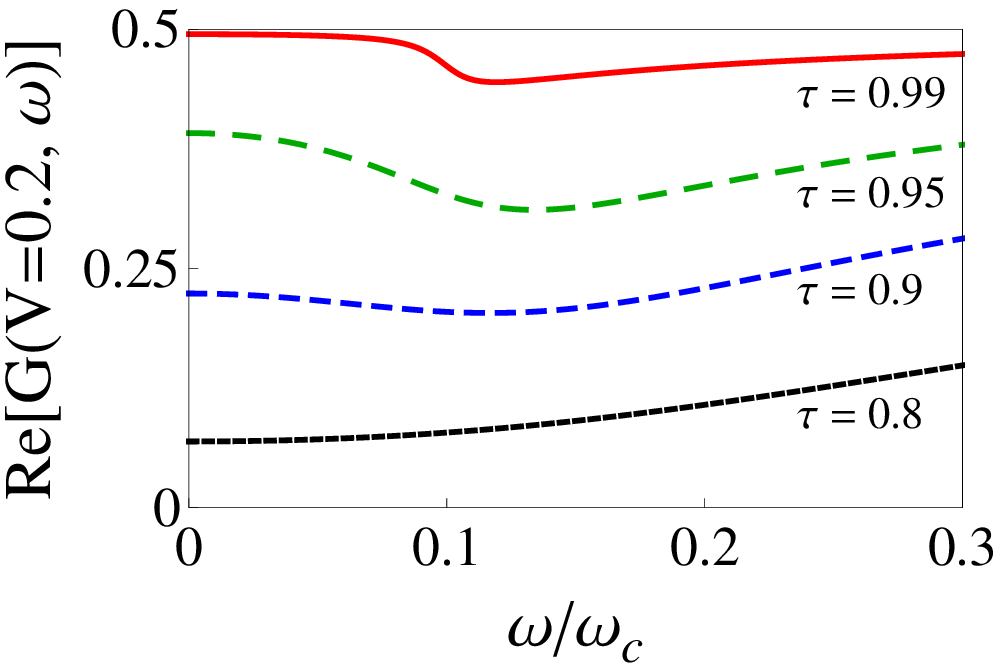}
\includegraphics[width=4.2cm]{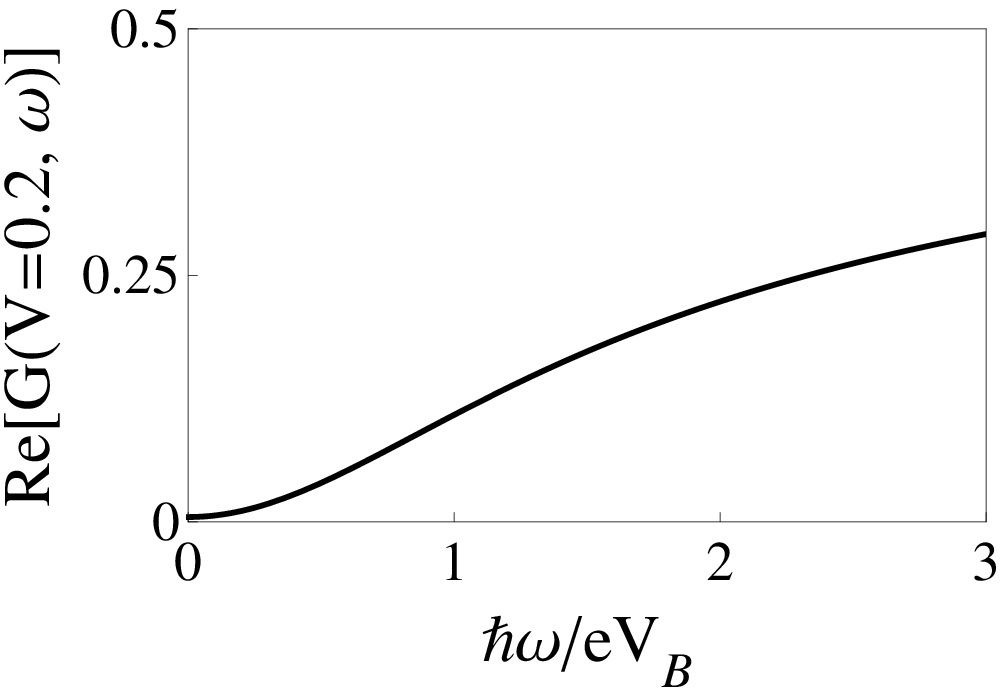}
\caption{Left panel: FF conductance, in units of $e^2/h$, as a function of $\omega/\omega_c$, for different values of $\tau$, at $eV/\hbar\omega_c=0.2$ and $k_BT/\hbar\omega_c=0.001$. Right panel: All the curves of the left graphic scale to a single one when one considers their variation with $\hbar\omega/eV_B$. We take $V/V_B=0.2$ and $k_BT/eV_B=0.001$. \label{figure6}}
\end{center}
\end{figure}

{\it Low temperature behavior}. At $T=0$, using Eqs.~(\ref{eq_G_I}) and (\ref{eq_current_T_0}), the FF conductance reads as:
\begin{eqnarray}
\mathrm{Re}[G(V,\omega)]=\frac{G_q}{2}\left[1-\frac{eV_B}{4\hbar\omega}\sum_\pm\mathrm{arctan}\left(\frac{2\hbar\omega\pm e V}{eV_B}\right)\right]~.\nonumber\\
\end{eqnarray}
For $\tau=1$, it reduces to the perfect conductance -- equal to $G_q/2$ -- whereas for $\tau< 1$, it acquires a frequency dependence as shown on Fig.~\ref{figure6}. Since $\mathrm{Re}[G(V,\omega)]$ is an even function of $\omega$, it is plotted at positive frequencies only. This result is interesting in the sense that the FF conductance has a non monotonous behavior. It exhibits a minimum (see the left panel of Fig.~\ref{figure6}) at a frequency depending on both  $V$ and $V_B$ which is fixed by the cancellation of the derivative $\partial_\omega\mathrm{Re}[G(V,\omega)]$. When $\tau$ decreases, this non monotonous behavior disappears, and the FF conductance increases regularly with frequency. All these curves scale to the same one when one considers their variation with respect to $\hbar\omega/eV_B$, at fixed $V/V_B$ and $k_BT/eV_B$ (see right panel of Fig.~\ref{figure6}). Notice that because of the finite value of the voltage, the FF conductance acquires a non-zero value -- even though small -- at zero frequency. This has been checked by zooming around $\omega=0$ (not shown).

We turn now our attention to the FF noise. At $T=0$, the integral in Eq.~(\ref{eq_noise}) can be performed analytically and the FF non-symmetrized noise reads:
\begin{eqnarray}\label{eq_noise_T_0}
S(V,\omega)&=&G_qeV_B\mathcal{F}\left(\frac{V}{V_B},\frac{\hbar\omega}{eV_B}\right)~,
\end{eqnarray}
where the dimensionless function $\mathcal{F}$ is given by:
\begin{eqnarray}\label{eq_fct_F}
\mathcal{F}(\tilde V,\tilde\omega)&=&-\tilde\omega\Theta(-\tilde\omega)+\frac{1}{8}\sum_{\pm}\Bigg[\pm\Theta(-\tilde\omega\pm \tilde V)\mathrm{arctan}(\tilde V)\nonumber\\
&&+\big[3\Theta(-\tilde\omega)-\Theta(-\tilde\omega\mp \tilde V)\big]\mathrm{arctan}\left({2\tilde\omega\pm \tilde V}\right)\Bigg]\nonumber\\
&&+\frac{1}{8\tilde\omega}\sum_{\pm}\big[-\Theta(-\tilde\omega)+\Theta(-\tilde\omega\pm \tilde V)\big]\nonumber\\
&&
\times\Bigg[\mathrm{ln}\left(1+{\tilde V^2}\right)-\mathrm{ln}\left(1+{(2\tilde\omega\mp \tilde V)^2}\right)\Bigg]~.
\end{eqnarray}
Here $\Theta$ is the Heaviside function, $\tilde V=V/V_B$ and $\tilde\omega=\hbar\omega/eV_B$. Notice that the FF noise is an even function of $V$. Because of this parity, we will specify only to positive values of the voltage. We recall that the FF non-symmetrized noise at positive (negative) frequencies corresponds to emission (absorption) noise.\cite{billangeon06}

An important fact we observe is that the emitted noise vanishes exactly above $eV$. In the perturbative regime, one usually expects that the emitted noise vanishes above $eKV$.\cite{bena07} Indeed, it has been shown universally for a weak tunneling junction (between strongly correlated systems in arbitrary dimensions and with possible coupling to an environment) that the FF noise associated to tunneling vanishes above $qV$, where $q$ is the transferred charge. \cite{safi09} Nevertheless, when higher order processes with respect to tunneling are taken into account, one expects the implication of a many-body complicated process, and energy conservation at the level of one particle can not be used any more. 
In strongly correlated systems or in a conductor coupled to an environment, we are not aware of any non-perturbative statement about the existence of a value of the frequency $\omega$ over which the emitted noise vanishes. However, we obtain an exact result here: the emitted noise vanishes strictly at frequencies $\hbar\omega>eV$, whatever the values of $V$ and $V_B$ are. This is due to the Bose-Einstein distribution functions of Eq.~(\ref{eq_noise_bis}) which reduce, in the zero temperature limit, to the Heaviside functions of Eq.~(\ref{eq_fct_F}). Their presence is related to the photon exchange between the conductor and the environment corresponding to electron-hole type excitations. One can think of two possible scenarios: (i) The new fermions are independent. For non-interacting electrons, scattering theory predicts that the emitted noise vanishes above $eV$. Even though the FF noise in our case does not obey the same relation with respect to the transmission amplitude $t(\omega)$, one can imagine that a similar fact holds. (ii)~The second scheme is that the system cannot emit at higher frequencies compared to what the generator can afford, and not compared to the voltage across the conductor $V/2$ if it was perfect. If this is plausible, it opens the question as to whether such a result is universal for the TLL with an arbitrary value of the parameter $K$, thus for other values of the resistance $R$. Equivalently, for $\hbar\omega<-eV$, we have $S(V,-\omega)=0$, and Eq.~(\ref{def_G}) reduces to:\cite{safi09}
\begin{eqnarray}
S(V,\omega<-eV/\hbar)=-2\hbar\omega\mathrm{Re}[G(V,\omega)]~,
\end{eqnarray}
which means that at negative frequencies below $-eV/\hbar$ and zero temperature, the absorption FF noise is governed by the FF conductance.

It is also interesting to express the FF non-symmetrized excess noise which is often measured experimentally, both because zero-point fluctuations are not easy to measure (apart from a pioneering work in Ref.~\onlinecite{basset10}), and in order to subtract undesirable sources of noise. It is given by:
\begin{eqnarray}
\Delta S(V,\omega)&=&S(V,\omega)-S(V=0,\omega)~.
\end{eqnarray}
At zero temperature, it reads: 
\begin{eqnarray}
\Delta S(V,\omega)=G_qeV_B[\mathcal{F}(\tilde V,\tilde\omega)-\mathcal{F}(0,\tilde\omega)]~, 
\end{eqnarray}
where the function $\mathcal{F}$ is given by Eq.~(\ref{eq_fct_F}).

\begin{figure}[!h]
\begin{center}
\includegraphics[width=4.2cm]{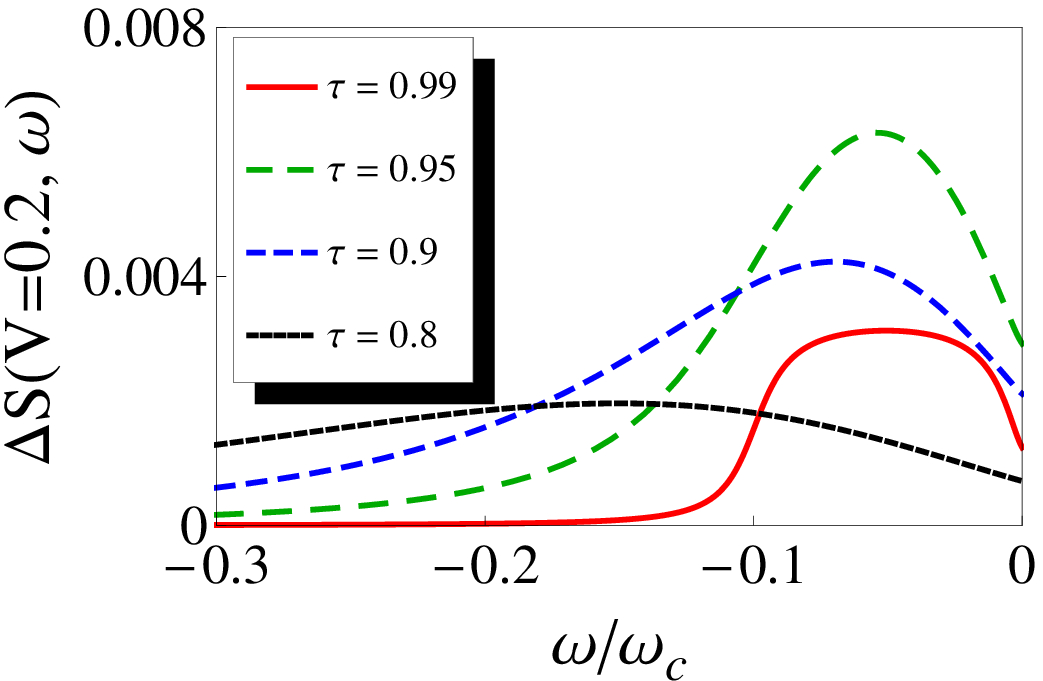}
\includegraphics[width=4.2cm]{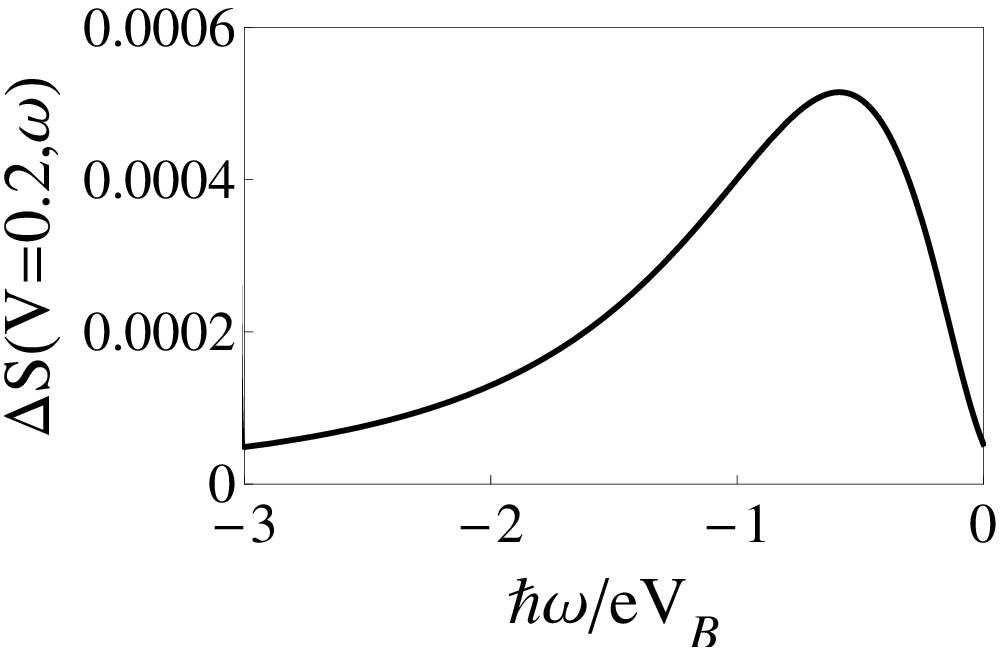}
\caption{Left panel: Non-symmetrized excess noise at negative frequency, in units of $e^2\omega_c$, as a function of $\omega/\omega_c$, for different values of $\tau$, at $eV/\hbar\omega_c=0.2$ and $k_BT/\hbar\omega_c=0.001$. Right panel: Non-symmetrized excess noise at negative frequency, in units of $e^2V_B/\hbar$, as a function of $\hbar\omega/eV_B$, at $V/V_B=0.2$ and $k_BT/eV_B=0.001$.\label{figure7}}
\end{center}
\end{figure}

\begin{figure}[!h]
\begin{center}
\includegraphics[width=4.2cm]{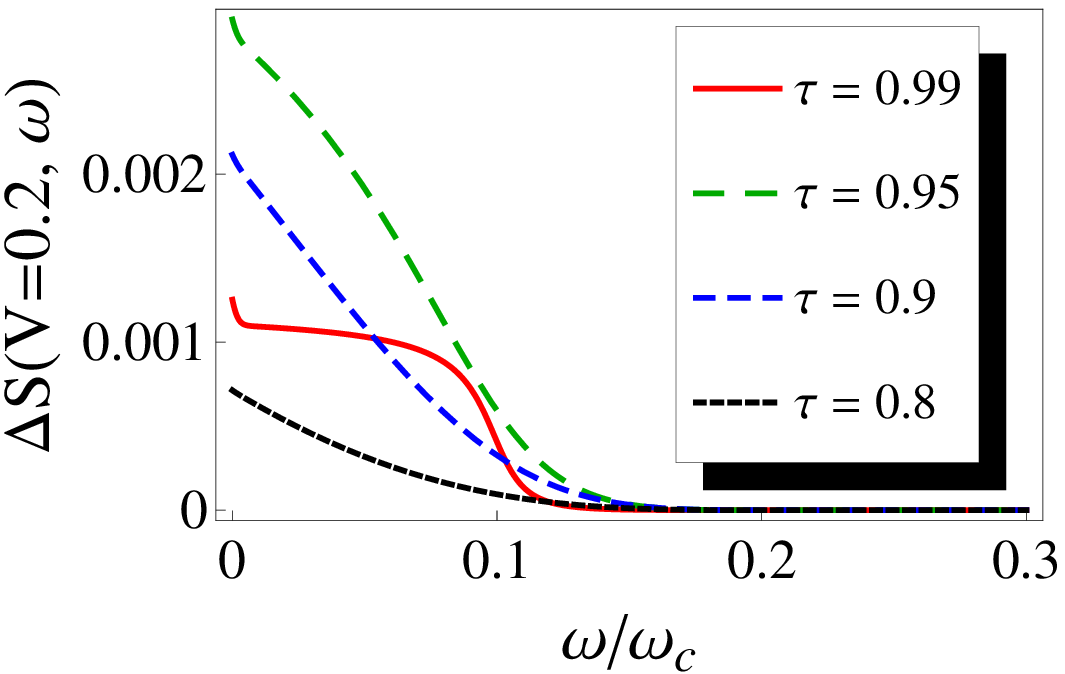}
\includegraphics[width=4.2cm]{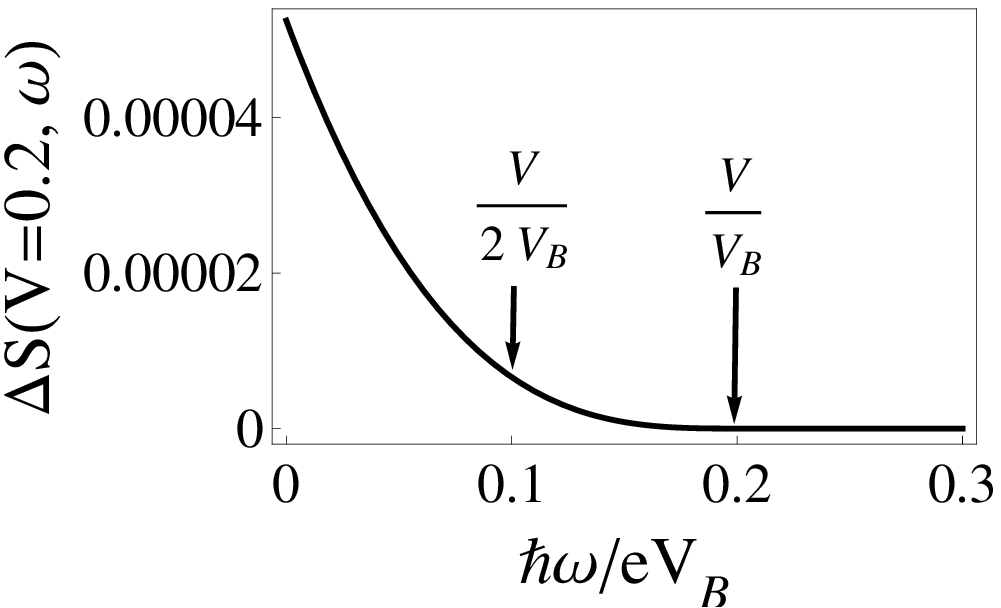}
\caption{Left panel: Non-symmetrized excess noise at positive frequency, in units of $e^2\omega_c$, as a function of $\omega/\omega_c$, for different values of $\tau$, at $eV/\hbar\omega_c=0.2$ and $k_BT/\hbar\omega_c=0.001$. Right panel: Non-symmetrized excess noise at positive frequency, in units of $e^2V_B/\hbar$, as a function of $\hbar\omega/eV_B$, at $V/V_B=0.2$ and $k_BT/eV_B=0.001$.\label{figure8}}
\end{center}
\end{figure}

In the left panels of Figs.~\ref{figure7} and \ref{figure8} is plotted the excess noise at a very low temperature\cite{note_temp} at positive frequency and negative frequency respectively. The red curve corresponds to a value of $V_B\ll V$, i.e. to the WBS regime. In that case one sees, according to Eq.~(\ref{eq_fct_F}), that the dependence close to $\hbar\omega=\pm e^*V=\pm eV/2$ is due mainly to the $\arctan$ function: this explains the step whose width is controlled by $V_B$. The step is smoothed out at higher $V_B$ (i.e., lower $\tau$) in the green and blue dashed curves of Fig.~\ref{figure7}. However, the asymmetry between absorbed and emitted excess noise is always present whatever the value of the effective transmission is. This asymmetry has been shown to be related, in a universal way, to non-linearity, using Eq.~(\ref{def_G}).\cite{safi09} Here, the non-linearity is induced by the electromagnetic environment. All these curves at different $\tau$ scale to a unique one as it is shown in the right panels of Figs.~\ref{figure7} and \ref{figure8} where $\Delta S$ is plotted as a function of $\hbar\omega/eV_B$ when the ratios $V/V_B$ and $k_BT/eV_B$ are fixed. We have to draw attention to the fact that this requires us to change simultaneously all energy scales while $V_B$ changes.

\begin{figure}[!h]
\begin{center}
\includegraphics[width=4.2cm]{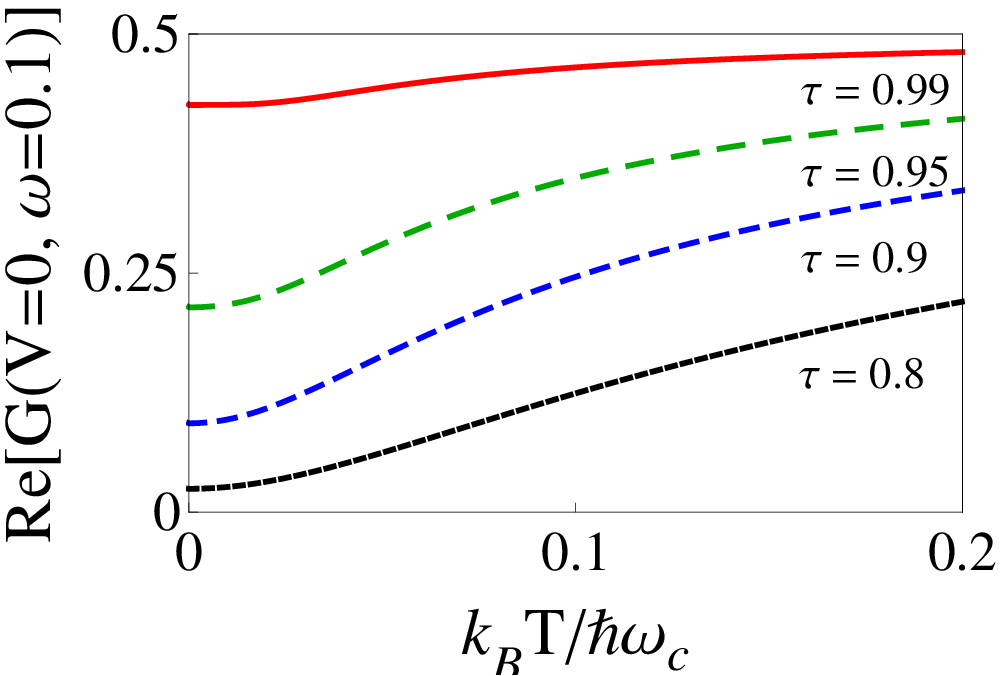}
\includegraphics[width=4.2cm]{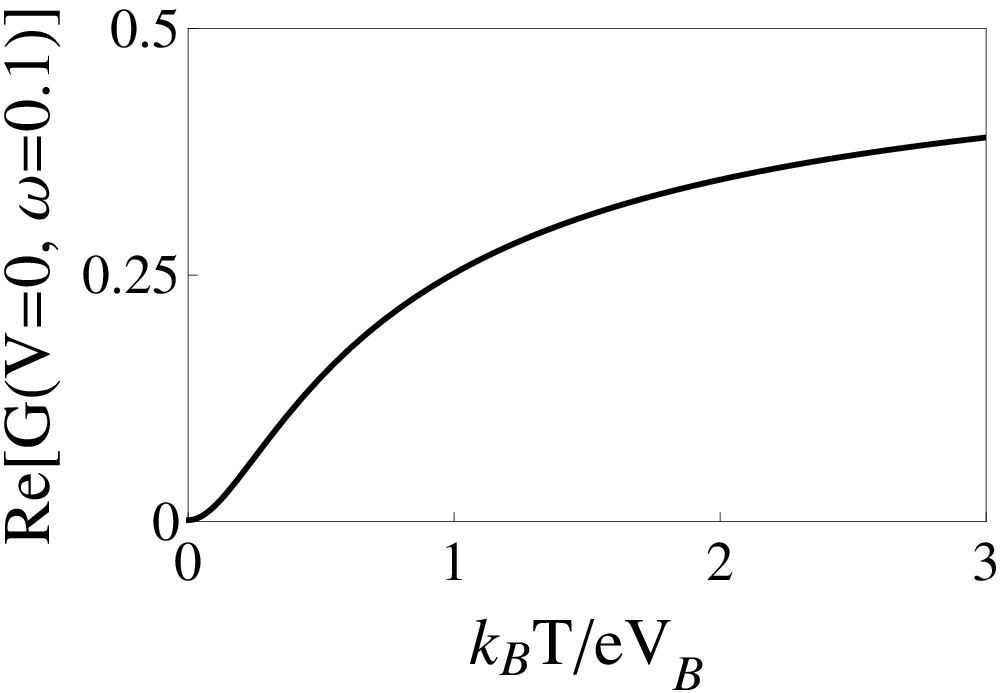}
\caption{Left panel: FF conductance, in units of $e^2/h$, as a function of $k_BT/\hbar\omega_c$, for different values of $\tau$, at $V=0$ and $\omega/\omega_c=0.1$. Right panel: All the curves of the left graphic scale to a single one when one considers their variations with respect to $k_BT/eV_B$. We take $V=0$ and $\hbar\omega/eV_B=0.1$.\label{figure9}}
\end{center}
\end{figure}

\begin{figure}[!h]
\begin{center}
\includegraphics[width=4.2cm]{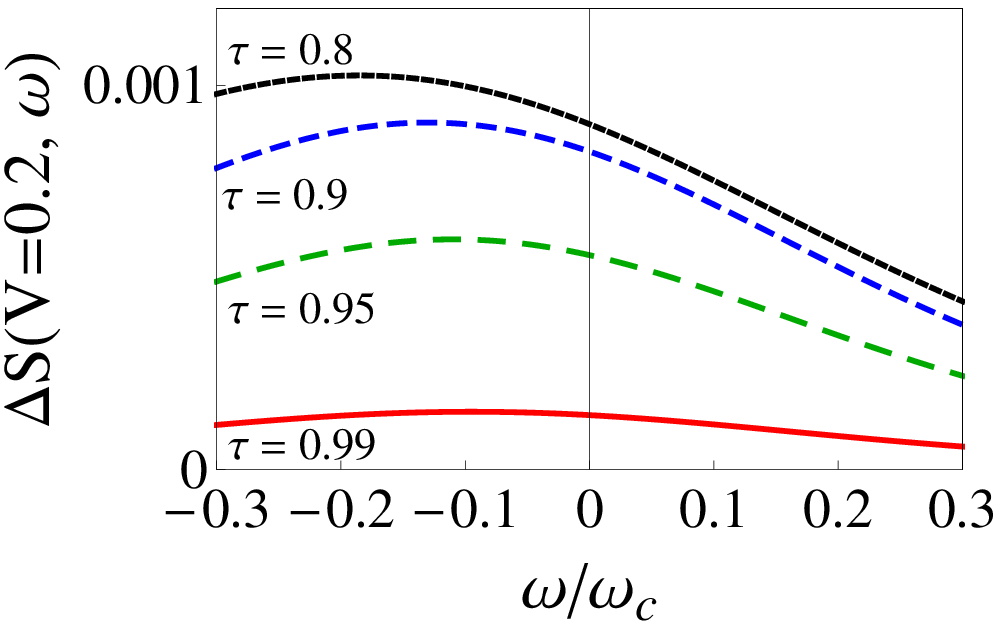}
\includegraphics[width=4.2cm]{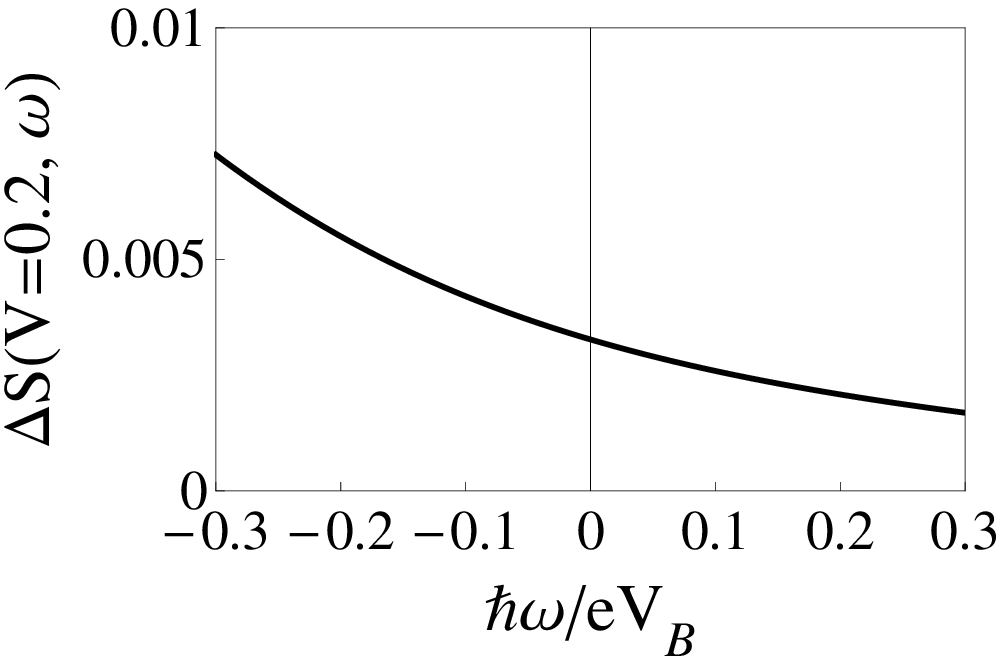}
\caption{Left panel: Non-symmetrized excess noise, in units of $e^2\omega_c$, as a function of $\omega/\omega_c$, for different values of $\tau$. We use the values $eV/\hbar\omega_c=0.2$, and $k_BT/\hbar\omega_c=0.2$. Right panel: Non-symmetrized excess noise, in units of $e^2V_B/\hbar$, as a function of $\hbar\omega/eV_B$, at $V/V_B=0.2$ and $k_BT/eV_B=0.2$.\label{figure10}}
\end{center}
\end{figure}

{\it Intermediate temperature behavior}. Next, we look at the effect of the temperature on the FF conductance and the FF noise. The left panel of Fig.~\ref{figure9} shows that the FF conductance increases monotonously with temperature. Again, all the curves scale to a single one when one considers the variation with respect to $k_BT/eV_B$ at fixed $\hbar\omega/eV_B$ (see the right panel of Fig.~\ref{figure9}). The FF excess noise at intermediate temperatures is plotted on the left panel of Fig.~\ref{figure10}. However, the asymmetry of the FF excess noise is still visible in that regime, and we observe its enhancement when the effective transmission decreases. Again, all the curves scale to a single one when one considers the variation with $\hbar\omega/eV_B$ at fixed values of $V/V_B$ and $k_BT/eV_B$ (see the right panel of Fig.~\ref{figure10}).

{\it High temperature behavior}. At high temperatures compared to voltage, $k_BT\gg eV$, we have checked, using Eqs.~(\ref{eq_noise_bis}) and (\ref{eq_G_I}), that the FF non-symmetrized noise is related to the FF conductance and becomes voltage independent in that limit according to the fluctuation-dissipation theorem:
\begin{eqnarray}
S(V\ll k_BT/e,\omega)=2\hbar\omega N(\hbar\omega)Re[G(V\ll k_BT/e,\omega)]~.\nonumber\\
\end{eqnarray}


    \section{Conclusion}

    In this paper, we have studied both stationary and time-dependent transport properties of a well-transmitting one-channel conductor embedded in an Ohmic environment with a quantum of resistance $R_q=h/e^2$. We have taken advantage of the mapping of this problem to a TLL with a particular value of the interaction parameter $K=1/2$. This has allowed us to obtain results which are non-perturbative with respect to the resistance of the environment $R$ (here being equal to $R_q$, see Eq.~(\ref{key})), and which describes all regimes of voltages, frequencies and temperatures below the cutoff $\omega_c$. The results are controlled by a unique scaling and non-universal parameter $V_B$ which depends on both the properties of the conductor and the environment. While the dc properties can as well be derived exactly for other values of $R$ (analytically at zero temperature, otherwise one would need numerical methods), using the Bethe-Ansatz solution for the impurity problem in the TLL, the particular choice of $R=R_q$ has allowed us to obtain the first non-perturbative results for both the FF noise and the FF conductance. For the latter, we have used its universal relation to the asymmetry of the emission and absorption noise, via an out-of-equilibrium fluctuation-dissipation theorem type formula (see Eq.~(\ref{def_G})) derived in Ref.~\onlinecite{safijoyez}.

We have first recalled the results for the dc transport obtained in a TLL with parameter $K=1/2$  in order to apply them to our present problem and to discuss  them in more details. The key procedure is based on refermionization, i.e., casting the non-trivial strong correlation effects into new chiral fermions with an energy-dependent transmission coefficient ${\mathcal T}(\omega)$. This cannot be obtained by any ``adiabatic'' evolution from the effective transmission $\tau$ (which explains our choice for a different notation). The dc current through the mesoscopic conductor takes a form similar to that within the scattering theory, applied to the new fermions: it is expressed as the integral over energy of ${\mathcal T}(\omega)$ times the difference of the Fermi-Dirac distribution functions of the left and right reservoirs. Even for a good effective transmission $\tau$, one recovers the DCB at energies below the voltage $V_B$. The reduction to the dc current has a power law with different exponents when voltages are high or low compared to $V_B$, corresponding to the WBS and SBS regimes. In particular, in the latter limit, one recovers the behavior predicted within the $P(E)$ theory for $V\ll V_B$ for a weakly transmitting conductor, even though we consider a well transmitting one here. In these two opposite limits, the ZF noise becomes Poissonian, with a Fano factor respectively given by $e/2$ or $e$. The former renormalization is related to the conductance of the conductor in the perfect limit, being in series with the environmental quantum resistance.

The ZF noise obeys as well the same expression in terms of ${\mathcal T}(\omega)$ as within the scattering approach. Nevertheless, this does not hold  for the FF noise even though it can be expressed in terms of the transmission amplitude $t(\omega)$ : this is a surprising and crucial result of our paper. Another interesting fact is that the emitted FF noise vanishes strictly above $eV$ at zero temperature, a fact common to non-interacting electrons, which is not obvious to expect within the underlying strongly correlated system.
We have shown that the FF conductance does not obey the scattering approach formulation for the new fermions. Rather, it obeys a simple relation that involves dc currents. This kind of relation was previously shown for tunneling barriers within the Tien-Gordon theory, or in the FQHE at simple fillings.\cite{bena07} Recently it has been extended to arbitrary filling factors, and even more has been shown to be universal either for tunneling barriers in arbitrary dimensions, as well as weak barriers in one dimension.\cite{safi10} It is quite remarkable that such a relation extends to arbitrary regimes within the TLL model at $K=1/2$.  It turns out that the FF conductance has a non-monotonous behavior with respect to frequency, having a minimum at a frequency that depends both on the applied voltage and the scaling voltage $V_B$, and the value of which is reduced when the effective transmission $\tau$ decreases. 

    We have shown that both the FF conductance and the FF noise have a universal behavior at different $\tau$ when voltages, temperatures and frequencies are all divided by the same scaling voltage $V_B$. This extends the result obtained for the differential conductance, valid for arbitrary values of the parameter $K$ -- thus of the resistance $R$ in our problem -- where we expect to get the same scaling behavior for time-dependent transport as well.

    The coherent conductor connected to an Ohmic environment  offers a unique framework to realize a TLL with a tunable parameter $K$, and a unique possibility to realize $K=1/2$. The present experiments on that direction are very promising.\cite{pierre11} Beyond this issue, our work provides benchmark results for time-dependent
     transport in various fundamental problems:  the DCB phenomena, other
    strongly correlated systems where the special value $K=1/2$ has been
    studied fully,\cite{komnik} and more generally those where one  solves the
    interacting problem in terms of new independent particles, such as
    within the Bethe-Ansatz methods for the impurity problem in TLL.\cite{fendley95} We have
    highlighted novel counter-intuitive facts, in particular, we have
    shown that one can not systematically apply the scattering approach
   to those new independent particles.

    \acknowledgments

    I.S. is grateful to A.~Anthore, S.~J\'ezouin, F.D.~Parmentier, F.~Pierre  for their insightful remarks and to S.~Carr, B.~Dou\c cot, D.~Est\`eve, C.~Glattli, P.~Joyez, H.~Saleur, P.~Simon, J.R.~Souquet, E.~Sukhorukov and A.~Zaikin for fruitful discussions; she acknowledges financial support of ANR under contract ANR-09-BLAN-0199-01. A.C. and R.Z. thank T.~Martin for preliminary discussions about the refermionization procedure.


\appendix

\section{Calculation of the current}\label{app_current}

\begin{figure}[!ht]
\begin{center}
\includegraphics[width=8cm]{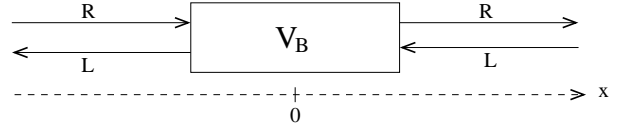}
\caption{Schematic representation of the left and right propagating channels in the reservoirs. $eV_B=2\hbar\omega_Fv_B^2$ is the energy scale that characterizes backscattering of electrons at the conductor position $x=0$.\label{figure11}}
\end{center}
\end{figure}

The mapping of Ref.~\onlinecite{safi04} applies to the effective action once degrees of freedom apart from those at $x=0$ are integrated out. It is more convenient, for the present computation, to consider the extended TLL Hamiltonian over space coordinate
$x$, with an impurity located at $x=0$ (see Fig.~\ref{figure11}):
\begin{eqnarray}\label{H_annexe}
\mathcal{H}&=&\frac{\hbar v_F}{4\pi} \int dx[(\partial_x\phi(x,t))^2+(\partial_x\tilde\phi(x,t))^2]\nonumber\\
&&+\sqrt{\frac{\hbar\omega_FeV_B}{32\pi^3}}e^{i\phi(0,t)-ieVt/2\hbar}+h.c.~,
\end{eqnarray}
where the bosonic fields $\phi$ and $\tilde\phi$ are related
to the initial bosonic fields describing right (R) and left (L) movers through:
\begin{eqnarray}
\phi(x,t)&=&\frac{1}{\sqrt{2}}\left(\phi_R(x,t)+\phi_L(x,t)\right)~,\\
\tilde\phi(x,t)&=&\frac{1}{\sqrt{2}}\left(\phi_R(x,t)-\phi_L(x,t)\right)~,
\end{eqnarray}
where $\psi_{L,R}=\eta_{L,R} e^{i\phi_{L,R}(x,t)}/\sqrt{2\pi a}$ are the initial fermionic operators.  $a$ is the distance cutoff of the TLL theory. The Klein factors $\eta_{L,R}$ allow to have the proper commutation relation for the fermionic operators. Notice that in Eq.~(\ref{hamilton}), the bosonic field refers to $\phi(t)\equiv\phi(0,t)$.

The average current through the conductor is given by:\cite{egger98}
\begin{eqnarray}\label{A1}
I=ev_F\langle\hat\rho_R-\hat\rho_L\rangle~,
\end{eqnarray}
where $v_F$ is the Fermi velocity. The operators $\hat\rho_{R}$ and $\hat\rho_{L}$ refer to the densities of right and left moving electrons in TLL reservoirs (see Fig.~\ref{figure11}). They are related to the new fermionic operators introduced in the refermionization procedure through the relations:
\begin{eqnarray}\label{A2}
\hat\rho_R(x,t)&=&\frac{\tilde\psi^\dag(x,t)\tilde\psi(x,t)+\psi^\dag(x,t)\psi(x,t)}{2}~,\\\label{A3}
\hat\rho_L(x,t)&=&\frac{\tilde\psi^\dag(-x,t)\tilde\psi(-x,t)-\psi^\dag(-x,t)\psi(-x,t)}{2}~,\nonumber\\
\end{eqnarray}
where $\psi(x,t)=\eta e^{i\phi(x,t)}/\sqrt{2\pi a}$ and $\tilde\psi(x,t)=\tilde\eta e^{i\tilde\phi(x,t)}/\sqrt{2\pi a}$ are the new fermionic fields associated to the bosonic fields $\phi$ and $\tilde\phi$: $\psi(x,t)$ is affected by backscattering and by the applied voltage (see Eq.~(\ref{H_annexe})), whereas $\tilde\psi(x,t)$ is neither affected by backscattering, nor by the applied voltage. In Eq.~(\ref{A1}), the average value is defined as the average over the scattering state\cite{chamon96} minus the average at equilibrium (i.e., $V=0$), since we have to take the normal order\cite{vondelft98} in the products of operators that appear in Eqs.~(\ref{A2}) and (\ref{A3}). Thus, we immediately conclude that $\langle\tilde\psi^\dag(x,t)\tilde\psi(x,t)\rangle=0$. The Klein factors $\eta$ and $\tilde\eta$ allow us to have the proper commutation relation for the new fermionic operators.

From Eqs.~(\ref{A2}) and (\ref{A3}), we understand that $\hat\rho_R$ and $\hat\rho_L$ depend on the position $x$, however, the average of the difference $\langle\hat\rho_R-\hat\rho_L\rangle$, which appears in the current, does not depend on the position, since the average of the total current is a conserved quantity (it is the reason why we do not keep the $x$ dependency in the current).

Using this Hamiltonian, one can write the equations of motion for the fields. The solution for the fermionic field $\tilde\psi$, associated with the free bosonic field, is that of free propagating electrons, whereas the solution for $\psi$ is given by\cite{chamon96} (from now, we take $x>0$):
\begin{eqnarray}\label{sol1}
\psi(-x,t)&=&\frac{1}{\sqrt{2\pi a}\omega_F}\int_{-\infty}^{\infty} a_\omega e^{-i(\omega+\frac{eV}{2\hbar})\frac{x}{v_F}-i\omega t}d\omega~,\nonumber\\\\\label{sol2}
\psi(x,t)&=&\frac{1}{\sqrt{2\pi a}\omega_F}\int_{-\infty}^{\infty} b_\omega e^{i(\omega+\frac{eV}{2\hbar})\frac{x}{v_F}-i\omega t}d\omega~,\nonumber\\
\end{eqnarray}
where the operator $b_\omega=t(\omega)a_\omega+r(\omega)a^\dag_{-\omega}$ is a combination of the annihilation and creation operators, $a^\dag_\omega$ and $a_\omega$, which obey the commutation relation $\{a_\omega,a^\dag_{\omega'}\}=\delta_{\omega,\omega'}$. The transmission and reflection amplitudes read as:
\begin{eqnarray}
t(\omega)&=&\frac{2\hbar\omega}{2\hbar\omega+ieV_B}~,\\
r(\omega)&=&\frac{ieV_B}{2\hbar\omega+ieV_B}~.
\end{eqnarray}

With the help of the solutions given by Eqs.~(\ref{sol1}) and (\ref{sol2}), one can calculate the average over the product of the two fermionic fields:
\begin{eqnarray}\label{av1}
&&\langle\psi^\dag(-x,t)\psi(-x,t)\rangle=\frac{1}{2\pi v_F}\int_{-\infty}^{\infty} d\omega\nonumber\\
&&\times\left[f\left(\hbar\omega-eV/2\right)-f\left(\hbar\omega\right)\right]~,
\end{eqnarray}
and,
\begin{eqnarray}\label{av2}
&&\langle\psi^\dag(x,t)\psi(x,t)\rangle=\frac{1}{2\pi v_F}\int_{-\infty}^{\infty} d\omega\nonumber\\
&&\times\left(2\mathcal{T}(\omega)-1\right)\left[f\left(\hbar\omega-eV/2\right)-f\left(\hbar\omega\right)\right]~,
\end{eqnarray}
where $\mathcal{T}=t^*t$. We have used the following average values: $\langle a_\omega a_{\omega'}\rangle=0$, and
$\langle a^\dag_\omega a_{\omega'}\rangle=f(\omega-eV/2)\delta_{\omega,\omega'}$, with $f$ is the Fermi-Dirac distribution function. Reporting Eqs.~(\ref{av1}) and (\ref{av2}) into Eqs.~(\ref{A1}) and (\ref{A2}), we finally obtain Eq.~(\ref{eq_current}).

\section{Noise calculation}\label{app_noise}

To calculate the FF non-symmetrized noise, we need to evaluate the following correlators ($x>0$):
\begin{eqnarray}
S_1&=&\int_{-\infty}^{\infty}dt e^{i\omega t}\langle\psi^\dag(-x,0)\psi(-x,0)\psi^\dag(-x,t)\psi(-x,t)\rangle~,\nonumber\\
S_2&=&\int_{-\infty}^{\infty}dt e^{i\omega t}\langle\psi^\dag(x,0)\psi(x,0)\psi^\dag(-x,t)\psi(-x,t)\rangle~,\nonumber\\
S_3&=&\int_{-\infty}^{\infty}dt e^{i\omega t}\langle\psi^\dag(-x,0)\psi(-x,0)\psi^\dag(x,t)\psi(x,t)\rangle~,\nonumber\\
S_4&=&\int_{-\infty}^{\infty}dt e^{i\omega t}\langle\psi^\dag(x,0)\psi(x,0)\psi^\dag(x,t)\psi(x,t)\rangle~.\nonumber\\
\end{eqnarray}

The determination of $S_1$ is rather simple since it involves contributions which are not affected by the backscattering:
\begin{eqnarray}
&&\langle\psi^\dag(-x,0)\psi(-x,0)\psi^\dag(-x,t)\psi(-x,t)\rangle=\frac{1}{4\pi^2a^2\omega_F^4}\nonumber\\
&&\times \int d\omega_1\int d\omega_2\int d\omega_3\int d\omega_4\langle a^\dag_{\omega_1}a_{\omega_2}a^\dag_{\omega_3}a_{\omega_4}\rangle\nonumber\\
&&\times e^{i(-\omega_1+\omega_2-\omega_3+\omega_4)\frac{x}{v_F}+i(\omega_3-\omega_4)t}~.
\end{eqnarray}

We use Wick's theorem to calculate the correlator $\langle a^\dag_{\omega_1}a_{\omega_2}a^\dag_{\omega_3}a_{\omega_4}\rangle$. After successive integrations over frequencies and times, we obtain :
\begin{eqnarray}
S_1&=&\frac{1}{2\pi v_F^2}\int_{-\infty}^{\infty}d\omega'f\left(\hbar\omega+\hbar\omega'-eV/2\right)\nonumber\\
&&\times\left[1-f\left(\hbar\omega'-eV/2\right)\right]~.
\end{eqnarray}

The calculation of $S_2$ and $S_3$ mix $a_\omega$ and $b_\omega$ operators since:
\begin{eqnarray}
&&\langle\psi^\dag(x,0)\psi(x,0)\psi^\dag(-x,t)\psi(-x,t)\rangle=\frac{1}{4\pi^2a^2\omega_F^4}\nonumber\\
&&\times \int d\omega_1\int d\omega_2\int d\omega_3\int d\omega_4\langle b^\dag_{\omega_1}b_{\omega_2}a^\dag_{\omega_3}a_{\omega_4}\rangle\nonumber\\
&&\times e^{i(-\omega_1+\omega_2-\omega_3+\omega_4)\frac{x}{v_F}+i(\omega_3-\omega_4)t}~,
\end{eqnarray}
and
\begin{eqnarray}
&&\langle\psi^\dag(-x,0)\psi(-x,0)\psi^\dag(x,t)\psi(x,t)\rangle=\frac{1}{4\pi^2a^2\omega_F^4}\nonumber\\
&&\times \int d\omega_1\int d\omega_2\int d\omega_3\int d\omega_4\langle a^\dag_{\omega_1}a_{\omega_2}b^\dag_{\omega_3}b_{\omega_4}\rangle\nonumber\\
&&\times e^{i(-\omega_1+\omega_2-\omega_3+\omega_4)\frac{x}{v_F}+i(\omega_3-\omega_4)t}~.
\end{eqnarray}

We obtain $S_2=S_3$ with:
\begin{eqnarray}
S_2&=&\frac{1}{2\pi v_F^2}\int_{-\infty}^{\infty}d\omega'f\left(\hbar\omega+\hbar\omega'-eV/2\right)\nonumber\\
&&\times\left[1-f\left(\hbar\omega'-eV/2\right)\right]\nonumber\\
&&\times \Big[t(\omega')t^*(\omega')t(\omega+\omega')t^*(\omega+\omega')\nonumber\\
&&-r(\omega')r^*(\omega')r(\omega+\omega')r^*(\omega+\omega')\Big]~.
\end{eqnarray}

Next, we calculate $S_4$ which involves $b_\omega$ and $b^\dag_\omega$ operators since:
\begin{eqnarray}
&&\langle\psi^\dag(x,0)\psi(x,0)\psi^\dag(x,t)\psi(x,t)\rangle=\frac{1}{4\pi^2a^2\omega_F^4}\nonumber\\
&&\times \int d\omega_1\int d\omega_2\int d\omega_3\int d\omega_4\langle b^\dag_{\omega_1}b_{\omega_2}b^\dag_{\omega_3}b_{\omega_4}\rangle\nonumber\\
&&\times e^{i(-\omega_1+\omega_2-\omega_3+\omega_4)\frac{x}{v_F}+i(\omega_3-\omega_4)t}~.
\end{eqnarray}

Applying Wick's theorem, we obtain:
\begin{eqnarray}
S_4&=&\frac{1}{2\pi v_F^2}\int_{-\infty}^{\infty}d\omega'\bigg\{f\left(\hbar\omega+\hbar\omega'-eV/2\right)
\nonumber\\
&&\times\left[1-f\left(\hbar\omega'-eV/2\right)\right]\nonumber\\
&&\times\Big[|t(\omega')|^2|t(\omega+\omega')|^2+|r(\omega')|^2|r(\omega+\omega')|^2\nonumber\\
&&-2t(\omega')t^*(\omega+\omega')r^*(\omega')r(\omega+\omega')\Big]\nonumber\\
&&+\sum_\pm f\left(\hbar\omega+\hbar\omega'\pm eV/2\right)\left(1-f\left(\hbar\omega'\mp eV/2\right)\right)\nonumber\\
&&\times \Big[|t(\omega')|^2|r(\omega+\omega')|^2\nonumber\\
&&+t(\omega')t^*(\omega+\omega')r^*(\omega')r(\omega+\omega')\Big]\bigg\}~.
\end{eqnarray}

Finally, the FF non-symmetrized noise of Eq.~(\ref{eq_noise}) is obtained from
$S(V,\omega)=e^2v_F^2[S_1+S_2+S_3+S_4]/4$, where we replace, everywhere it appears, $r(\omega)$ by $1-t(\omega)$.\\

\section{Comparison to the results obtained within the scattering theory}\label{app_comp}

In this Appendix, we compare the expressions of current, FF conductance and noise that we have established to those obtained in the framework of the scattering theory. 

It has been already noted  that the expressions of the dc current given by Eq.~(\ref{eq_current}) and the ZF noise are formally identical to those derived from the scattering approach (i.e., they are expressed as integrals of the transmission coefficient times Fermi Dirac distribution functions), even though the non-trivial many-body effects are encoded into this transmission. 
We show here that the FF conductance and the FF noise, even though expressed in terms of the transmission and reflection amplitudes, do not obey the relation derived within the scattering approach.

To make a comparison with the FF noise given in the literature,\cite{martin92,buttiker92,yang92,blanter00} we need to symmetrize the noise given by Eq.~(\ref{eq_noise}):
\begin{widetext}
\begin{eqnarray}\label{eq_sym_noise}
S(V,\omega)+{S(V,-\omega)}=\frac{e^2}{4\pi}\int_{-\infty}^{\infty}d\omega'\bigg(\Big[\mathcal{T}(\omega')\mathcal{T}(\omega+\omega')+|t(\omega')-t(\omega+\omega')|^2/4\Big]\Big[F_{++}(\omega,\omega')+F_{--}(\omega,\omega')\Big]\nonumber\\
+ \Big[\mathcal{T}(\omega')-\mathcal{T}(\omega')\mathcal{T}(\omega+\omega')-|t(\omega')-t(\omega+\omega')|^2/4\Big]\Big[F_{+-}(\omega,\omega')+F_{-+}(\omega,\omega')\Big]\bigg)~,
\end{eqnarray}
where $F_{ss'}(\omega,\omega')=\sum_\pm f(\hbar\omega'\pm\hbar\omega\pm seV/2)[1-f(\hbar\omega'\pm s'eV/2)]$. Using our notations, the symmetrized noise of  a coherent one-channel coherent conductor with an energy-dependent transmission $\mathcal{T}$ can be obtained within the scattering approach:\cite{buttiker92}
\begin{eqnarray}\label{eq_buttiker}
S_{\mathrm{scattering}}(V,\omega)+{S_{\mathrm{scattering}}(V,-\omega)}=\frac{e^2}{4\pi}\int_{-\infty}^{\infty}d\omega'\bigg(\big[\mathcal{T}(\omega')\mathcal{T}(\omega+\omega')+|t(\omega')-t(\omega+\omega')|^2\big]F_{++}(\omega,\omega')\nonumber\\
+\mathcal{T}(\omega')\mathcal{T}(\omega+\omega')F_{--}(\omega,\omega')+\mathcal{T}(\omega')(1-\mathcal{T}(\omega+\omega'))F_{+-}(\omega,\omega')+ \mathcal{T}(\omega+\omega')(1-\mathcal{T}(\omega'))F_{-+}(\omega,\omega')\bigg)~.
\end{eqnarray}
\end{widetext}

The main difference between Eqs.~(\ref{eq_sym_noise}) and (\ref{eq_buttiker}) resides into the factors in front of $F_{++}$ and $F_{--}$ which are identical in Eq.~(\ref{eq_sym_noise}) but not in Eq.~(\ref{eq_buttiker}). The same remark applies to the factors in front of $F_{+-}$ and $F_{-+}$. If one had a transmission coefficient $\mathcal{T}$ independent on energy, both expressions would have been identical.



\begin{thebibliography}{20}
\bibitem{note3}We use the notation $eV_B$ for the energy that characterizes the crossover between the WBS and SBS regimes instead of the notation $4k_BT_B$ used in Ref.~\onlinecite{fendley95}.
\bibitem{noteq}Notice that the effective charge $e^*$ corresponds here to $Ke$, but does not correspond to any physical entity within the DCB problem: it is related simply to the conductance in series of a perfect conductor, with a quantum of resistance, and the resistance $R$.
\bibitem{notecutoff}In general, in the TLL model, the cutoff procedure can contain more
than one cutoff energy, a tricky problem which we do not address fully
here. The mapping requires to cut the frequencies in the action at
$\omega_c$, thus the bosonic correlation functions are functions of
energy divided by $\hbar\omega_c$. Thus we adopt this unique cutoff
throughout the paper.
\bibitem{notesurtau} The low transmitting regime, $\tau\approx 0$, is not a good starting point for the theory. Indeed, in this limit, marginal process can dominate over the irrelevant tunneling process, and one can not obtain universal results. This is opposite to what happens in the good transmitting regime where the backscattering Hamiltonian in Eq.~(\ref{hamilton}) is relevant in the RG sense.
\bibitem{note4}From Eq.~(\ref{eq_current}), we directly see that the current is an odd function with voltage.
\bibitem{note5}From this relation, one can check easily that the real part of the FF conductance $\mathrm{Re}[G(V,\omega)]$ is an even function with voltage.
\bibitem{note_cond}At strictly zero-temperature, the integral in Eq.~(\ref{eq_G}) can be performed analytically and the derivative of the differential conductance according to $\tau$ can be obtained from Eq.~(\ref{diff_cond_T_0}), it reads: $\partial_\tau G(V,\omega=0)=4G_q\tau(1-\tau)e^2V^2\hbar^2\omega_c^2/(\tau^2e^2V^2+4(1-\tau)^2\hbar^2\omega_c^2)^2$. Since $eV\ll \hbar\omega_c$, we have $\partial_\tau G(V,\omega=0)\approx G_q\tau e^2V^2/4(1-\tau)^3\hbar^2\omega_c^2$ which can be quite large for $\tau\approx 1$.
\bibitem{note_temp}In Figs.~\ref{figure7} and \ref{figure8}, the curves are plotted at very low but small temperature in order to avoid artificial discontinuities.
\end{thebibliography}
\end{document}